\documentclass[11pt,a4paper,twocolumn, superscriptaddress]{revtex4-1}
\usepackage{amsthm}
\usepackage{graphicx}  
\usepackage{dcolumn}   
\usepackage{bm}        
\usepackage{amssymb}   
\usepackage{amsmath}
\usepackage{xcolor}
\usepackage{rotating}
\usepackage{xspace}
\usepackage{hyperref}

\newcommand{\cO}{{\cal O}}

\newcommand{\e}{\text{e}\,}
\newcommand{\bra}{\langle}
\newcommand{\ket}{\rangle}

\newcommand{\bear}{\begin{eqnarray}}
\newcommand{\eear}{\end{eqnarray}}

\renewcommand{\Re}{{\rm Re}}
\renewcommand{\Im}{{\rm Im}}

\def\be#1\ee{\begin{equation}#1\end{equation}}
\def\bea#1\eea{\begin{align}#1\end{align}}

\begin{document}

\title{
Complex Langevin and boundary terms 
}

\author{Manuel Scherzer}
\email{scherzer@thphys.uni-heidelberg.de}
\affiliation{ Institut f\"ur Theoretische Physik, Universit\"at Heidelberg, Heidelberg, Germany}
\author{Erhard Seiler}
\email{ehs@mpp.mpg.de}
\affiliation{Max-Planck-Institut f\"ur Physik (Werner-Heisenberg-Institut),  M{\"u}nchen, Germany}
\author{D\'enes Sexty}
\email{sexty@uni-wuppertal.de}
\affiliation{Institut f\"ur Theoretische Physik, Bergische Universit\"at Wuppertal, Wuppertal, Germany}
\affiliation{AS/JSC, Forschungszentrum J\"ulich, J\"ulich, Germany}
\author{Ion-Olimpiu Stamatescu}
\email{I.O.Stamatescu@thphys.uni-heidelberg.de}
\affiliation{ Institut f\"ur Theoretische Physik, Universit\"at Heidelberg, Heidelberg, Germany}

\date{\today}

\begin{abstract}
As is well-known the Complex 
Langevin (CL) method sometimes fails to converge or converges to the wrong 
limit. We identified one reason for this long ago: insufficient decay of 
the probability density either near infinity or near poles of the drift, 
leading to boundary terms that spoil the formal argument for correctness. 
To gain a deeper understanding of this phenomenon, we analyze the 
emergence of such boundary terms thoroughly in a simple model, where 
analytic results can be compared with numerics.  We also show how some 
simple modification stabilizes the CL process in such a way that it can
produce results agreeing with direct integration.  Besides 
explicitly demonstrating the 
connection between boundary terms and correct convergence our
analysis also suggests a correctness criterion which could be 
applied in realistic lattice simulations.

\end{abstract}
\maketitle
\newpage
\section{Introduction}
   
It has been known for a long time that the Complex Langevin (CL) method 
for simulating systems with complex action may fail by either not 
converging or by converging to the wrong limit. These failures were traced 
either to insufficient decay of the probability distribution in the 
complexified configuration space
\cite{Aarts:2009uq,Aarts:2011ax,Nishimura:2015pba} (at infinity or at 
poles of the drift force), or to failure of ergodicity 
\cite{Aarts:2017vrv,Seiler:2017wvd}. Recently Salcedo \cite{salc2} has 
formulated interesting criteria for failure that at first sight seem to be 
unrelated to the ones identified by us. The most interesting ones derive 
support properties of the equilibrium measure which are shown in 
these cases to be in conflict with the correct expectation values.

In this note we will focus on one such example and show explicitly that 
the problems are due to slow decay, leading to the appearance of boundary 
terms in an integration by parts, spoiling the formal proof of 
correctness. We stress that we are here concerned with the behavior at 
large non-compact dimensions. The effects of non-holomorphicity have been 
shown, e.g. in random matrix models \cite{Mollgaard:2013qra, 
Bloch:2017sex} to lead to wrong convergence and were specifically 
addressed in \cite{Aarts:2017vrv} both in simple models and in QCD.

Here we consider a complex density    
\be
\rho(x)=\exp(-S(x))\,,
\ee
periodic with period $2\pi$ and extending to an entire analytic function 
without zeroes.

The complex Langevin equation (CLE) in the form used here is 
\begin{align} 
dx=&K_x dt +dw,\notag\\ dy=&K_y dt\,, 
\label{cle2} 
\end{align} 
where $dw$ is the Wiener process normalized as \be \bra dw^2\ket= 2 dt\, 
\ee 
and the drift is given by 
\begin{align} 
K_x=& -{\rm Re}\,S'(x+iy) ,\cr K_y=& -{\rm Im}\,S'(x+iy)\,. 
\end{align} 
The long time asymptotic average of a generic observable $\cO$ is denoted 
by $\bra {\cO} \ket_\infty$; we say that the CL process yields 
correct results if this agrees with the `correct' expectation value of 
the same observable defined as 
\be 
\bra {\cO} \ket_c = \int dx {\cO}(x) \rho(x)\,, 
\ee
i.~e. 
\be 
\bra {\cO} \ket_\infty= \bra {\cO} \ket_c\,. 
\ee 
In \cite{Aarts:2009uq,Aarts:2011ax} correctness was derived from the 
consideration of CL expectation values at {\rm finite} Langevin time; it 
was shown that correctness is assured if a certain quantity 
$F_\cO(t,\tau)$ is independent of an interpolation parameter 
$\tau\in[0,t]$, i.~e. 
\be 
\label{interpol} 
\frac{\partial}{\partial \tau} F_\cO(t,\tau) = 0\,. 
\ee

Here $F_\cO(t,\tau)$ interpolates between the `correct' time evolution 
$F_\cO(t,t)=\bra\cO(t)\ket_0$ (defined in Section~\ref{boundary} and 
analyzed in Appendix~\ref{app2}) and the time evolution of the expectation 
of $\cO$ under the Langevin process $F_\cO(t,0)= \bra \cO\ket_t $. The 
key points are that (\ref{interpol}) implies

\be
\bra \cO\ket_t= \bra\cO(t)\ket_0 \quad \forall t > 0
\ee
and hence,
\be
\lim_{t\to\infty} F_\cO(t,t)=\bra\cO\ket_c\,.
\ee

The left hand side of (\ref{interpol}), by using integration by parts, is 
found to be equal to a boundary term; explicitly 
\be
\frac{\partial}{\partial \tau} F_\cO(t,\tau)=
\lim_{Y\to\infty}B_\cO(Y;t,\tau)\, ,
\ee
where
\bea
&B_\cO(Y;t,\tau)\equiv\\
&\int [K_y(x,Y)P(x,Y;t-\tau)\cO(x+iY;\tau)\notag\\&-
K_y(x,-Y)P(x,-Y;t-\tau)\cO(x-iY;\tau) ]dx\, ,
\label{boundt}
\eea

$P(x,y;t)$ is the time evolved probability density under the 
Langevin evolution and $\cO(t)\equiv\cO(z;t)$ is the $L_c$ 
evolved observable (see Section \ref{boundary} and appendix~\ref{app2}).

This form of the boundary term makes clear that correctness requires
sufficient decay of the product $K_y P\cO$ for all Langevin times 
$t$.

\section{The model}   

The model studied here is defined by the complex density
\be
\quad \rho = \frac{1}{Z(\beta)}\exp\left[-i\beta \cos(x)\right]\,
\label{model}
\ee
and has been studied already in 2007 by Stamatescu \cite{nucunotes}
and in 2008 by Berges and Sexty \cite{Berges:2007nr}. The `correct' 
expectation values of exponentials (`modes') are
\bea
\int dx \exp&(ikx)\rho(x)=\nonumber
\frac{I_k(-i\beta)}{I_0(-i\beta)}\\
=&(-i)^k\frac{J_k(\beta)}{J_0(\beta)}\neq 
0\,. \label{rhoexp}
\eea
It was found in \cite{nucunotes,Berges:2007nr} that the CL process does 
not reproduce the correct EV's which, however, can be regained by a 
certain reweighting procedure (with different observables requiring 
sometimes different reweightings).

The remarkable fact found by Salcedo \cite{salc2} is that the static
probability distribution $P(x,y)\equiv P(x,y;\infty)$ for this 
model can be written down explicitly by solving the time independent 
Fokker-Planck equation (FPE);
it is
\be
P(x,y)=\frac{1}{4\pi \cosh^2(y)}\,.
\label{P}
\ee
It is the only non-Gaussian example known to us for which a solution of 
the static FPE has been found in analytic form. Three features of 
this solution are remarkable:

(1) $P$ is independent of $x$,
   
(2) $P$ is independent of $\beta$,

(3) $P$ decays as $\exp(-2|y|)$ for large $|y|$; this decay is not
sufficient to make the integrals of the modes
\be
\exp(ik(x+iy)),\quad |k|\ge 2
\ee
absolutely convergent, in other words, already here we are faced with slow 
decay.

\subsection{Complex Langevin results}
   
But first let us demonstrate that (\ref{P}) is indeed the distribution
produced by running a CL simulation for a long time. The drift force is
\bea
&K_x=\Re\,\frac{\rho'}{\rho}=-\beta \cos(x)\sinh(y)\, \nonumber\\
&K_y=\Im\,\frac{\rho'}{\rho}=\beta\sin(x)\cosh(y)\,,
\label{drift}
\eea
for the Langevin process Eq.~(\ref{cle2}). 

\begin{figure}[ht]
\begin{center}
\includegraphics[width=1\columnwidth]{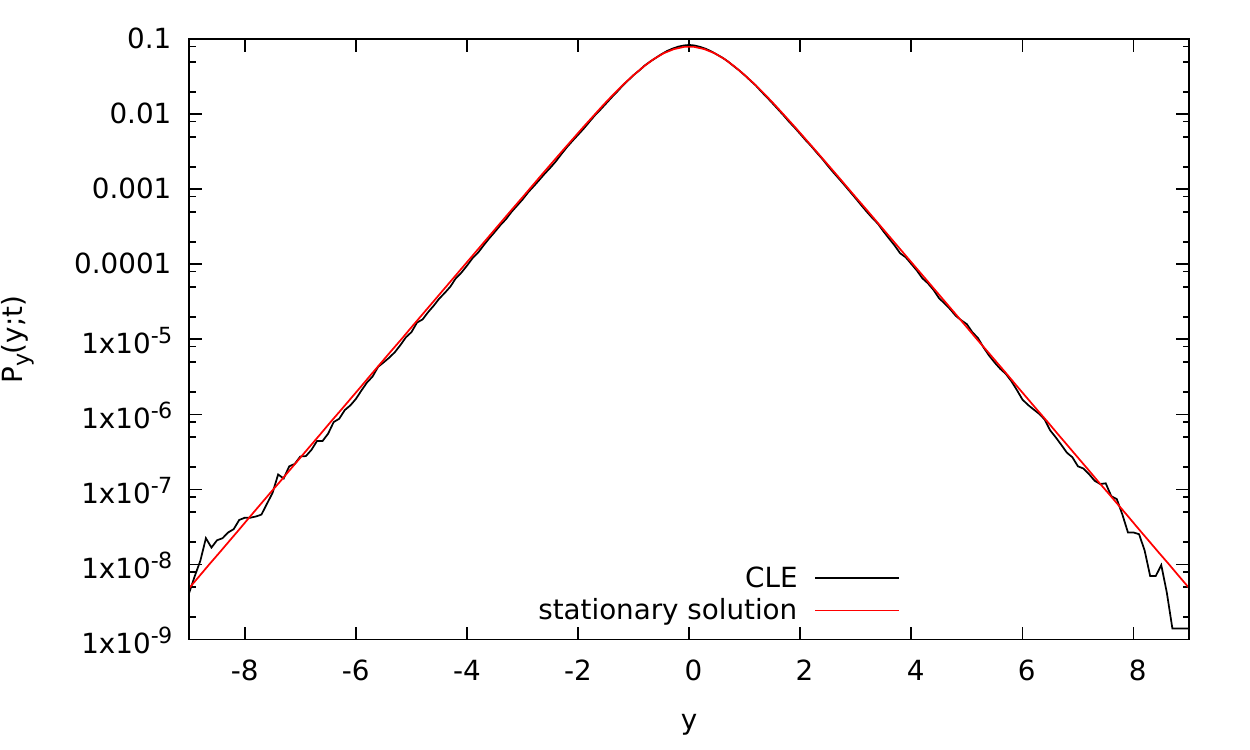}
\caption{Comparison of the analytic expression for the marginal 
distribution $P_y(y;t)$ (\ref{P}) (red) with the histogram of a CL 
simulation with $\beta=1$.}
\label{1dhisto}
\end{center}   
\end{figure}

In Fig.~\ref{1dhisto} we show the histogram of the converged marginal 
distribution $P_y(y;t)=\int dx P(x,y;t)$ in log scale for $\beta=1$, 
overlaid with the distribution (\ref{P}). The histogram is 
obtained from one long trajectory (Langevin time $t \approx$ 125000). The 
agreement over about 6 orders of magnitude is convincing. The distribution 
can also be seen to be independent of $x$, cf. also Fig.~\ref{marginalx}.

\begin{figure}[ht]
\begin{center}
\includegraphics[width=1\columnwidth]{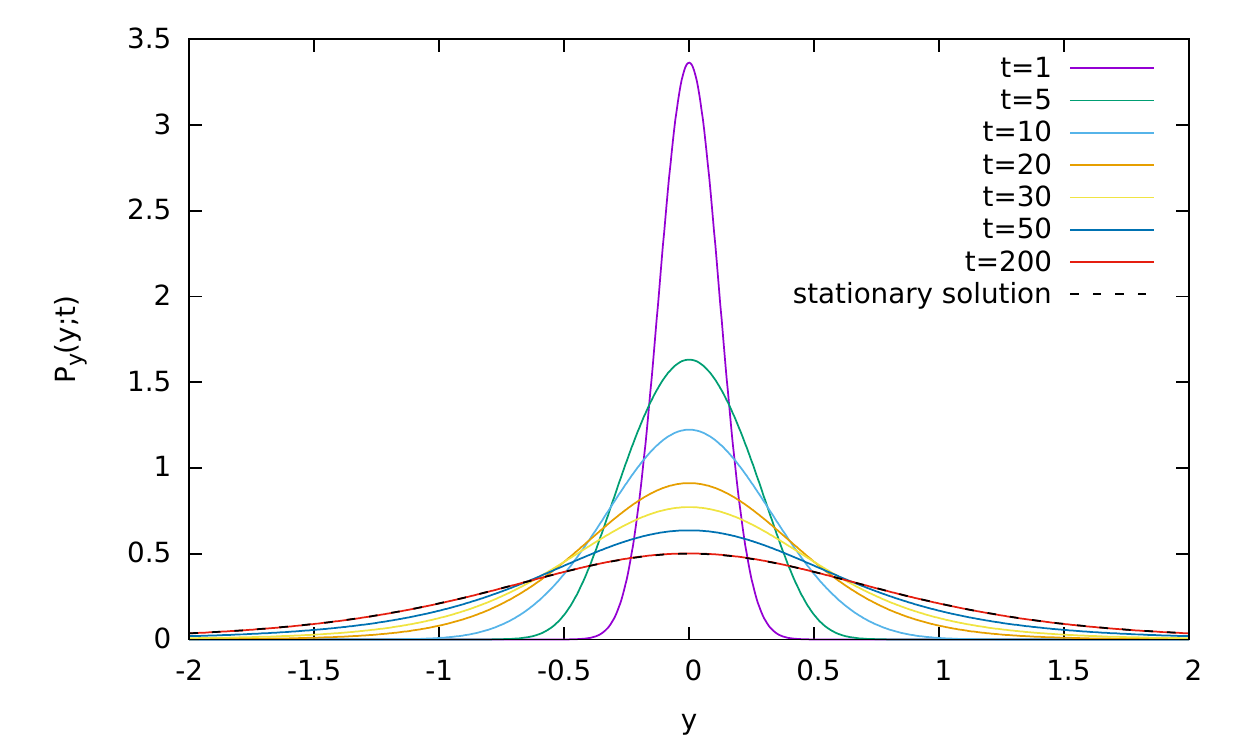}
\caption{The marginal distributions $P_y(y;t)$ obtained by numerically 
solving the FPE for $\beta=0.1$. The ordering of times corresponds to 
decreasing maxima; note that for $t=200$ no difference is visible between 
the FPE and the analytical solution.}
\label{marginal}
\end{center}
\end{figure}

We also show in Fig.~\ref{marginal} the histograms of $P_y(y;t)$ for
various shorter times and $\beta=0.1$, illustrating the convergence as 
$t\to\infty$.

As noted by Salcedo \cite{salc2}, it is obvious that the distribution $P$
(\ref{P}) cannot reproduce the correct expectation values Eq. \ref{rhoexp}
of the observables $\cO_k = \exp(ikx)$, because it is independent of $x$, 
entailing
\be
\int dx\,P(x,y) e^{\pm ikx}=0\,,
\ee
and its slow decay makes the expectation values of $\cO_k$ ill-defined for 
$|k|\ge 2$. In Table~\ref{salc1} we collect a few CL results, together 
with the exact expectation values determined by $\rho$ for $\beta=1$. The 
simulation used 100 independent trajectories with randomly chosen 
starting points on the real axis, running for a Langevin time of 
$t\approx 2500$, where measurements were taken after every time step, 
typically $5\times 10^{-6}$. The CL values for $|k|> 2$ are completely 
submerged by noise, as expected. For $k=\pm 2$ we find a value 
close to $1$. It should be remarked that the CL process for $k=2$ 
evaluates a conditionally convergent integral, so also the measuring 
schedule plays a role; for instance measuring after every time increment 
of $0.01$ yields very noisy results, consistent with both $0$ and $1$. 
Evaluating the second mode with a fixed cutoff in $y$, we find $0$.    

The Schwinger-Dyson equations (SDE) 
\be  
ik\bra e^{ikz}\ket+\frac{\beta}{2}\bra e^{i(k+1)z}\ket-
\frac{\beta}{2}\bra e^{i(k-1)z}\ket=0\,, 
\ee 
arising from the identity
\be
\int_{-\pi}^\pi\ \rho(x) \cO'(x)dx= - \int_{-\pi}^\pi \rho'(x) \cO(x)dx
\ee
would be satisfied for $k=0,\pm 1$ if the modes $\pm 1$ are $0$ and the 
modes $\pm 2$ are $1$, even though these values are not the `correct' 
ones.

\begin{table}[ht]
\begin{center}
\begin{tabular}{r||r |r| r| r}
$\bra \cO\ket$& $\bra \e^{ix}\ket$ & $\bra \e^{-ix}\ket$ & $\bra
\e^{2ix}\ket$ & $\bra\e^{-2ix}\ket$      \\
\hline
CL& 0.004(3) &  0.002(3) &1.027(22) & 1.001(20)\\
\hline
correct &-0.575081$i$ & -0.575081$i$ & -0.150162  & -0.150162 \\
\hline
$e^{t L_c}\cO$   &-0.575081$i$ & -0.575081$i$ &  -0.150162 &
-0.150162
\end{tabular}
\caption{CLE (real part, imaginary part negligible) and correct results 
for model (\ref{model}) with $\beta=1$. Last line: `correct evolution' for 
$t=20$ (see Appendix \ref{app2}).}
\label{salc1}
\end{center}
\end{table}
So the CL results, where they are defined, are incorrect, but mostly -- 
for $|k|\ge 3$, they are completely undefined due to uncontrollable
fluctuations.  

The last row in the table gives the correct results from the $L_c$
evolved observables, as will be explained in the next section. Notice that 
the correct results of course also satisfy the SDE, but these equations, 
having the structure of a two-step recursion, have a two-parameter 
family of solutions \cite{Berges:2006xc,Pehlevan:2007eq,Aarts:2011ax}.

\subsection{A puzzle}

The remaining question is: how can CL fail for the first mode, i.~e. 
observables $\cO_{\pm 1}\equiv \exp(\pm i(x+iy))$? $\cO_{\pm 1}P$ as well 
as $KP$ decay exponentially in $y$. 

Actually the densities of $\cO_{\pm 1}$ and $K$, if considered not 
as functions of $y$, but as functions of their actual value decay only 
power-like (see Eq.~\ref{sigma_anal}). Nagata et al \cite{Nagata:2016vkn} 
gave an argument that correctness requires exponential decay of the 
distribution of $K$ and checked their criterion successfully for various 
cases; so by this criterion correctness is not to be expected here, 
corroborating the criterion. We will, however, formulate a different 
criterion in Section~\ref{boundary}, which directly relates to the 
(non-)occurrence of boundary terms. 

The CL simulation produces for $\cO_{\pm 1}$ well converged, yet incorrect 
results, close to 0 (consistent with (\ref{P}) but inconsistent with 
(\ref{rhoexp})).

The resolution lies in the nonvanishing boundary terms arising in the time 
dependent expectation values and persisting for arbitrarily large times; 
this is the mechanism described in \cite{Aarts:2009uq,Aarts:2011ax}. In 
the following section we will analyze those boundary terms in detail.

\section{Boundary terms for finite Langevin time}
\label{boundary}

The formal argument for correctness \cite{Aarts:2009uq, Aarts:2011ax} is 
revisited in Appendix \ref{app1}. It requires the choice of an initial
distribution $P(x,y;0)$; in the following  we will choose for simplicity
\be
P(x,y;0)=\frac{1}{2\pi} \delta(y)\,.
\label{init}
\ee
The identity (\ref{interpol}) follows by integrating by parts, assuming 
that there are no boundary terms, and using the Cauchy-Riemann equations.

In order to check for the appearance of boundary terms as in   
(\ref{boundt}), we need the $L_c$ evolution of the observables 
(see below and Appendix \ref{app2}) and the time evolution of the 
probability density $P$ by solving the FPE with the initial condition 
(\ref{init}).
   
\subsection{Indirect evidence for boundary terms}
\label{indirect}

In \cite{Aarts:2011ax} we found numerically for a somewhat different model 
that the $L_c$ evolved observables $\cO(x+iy;t)$ grow in the $y$ 
direction as an iterated exponential. The same can be seen here, but we 
will not go into this. This growth makes the appearance of boundary terms 
already plausible.

In the following we show explicitly that Eq. (\ref{interpol}) is 
numerically satisfied for short times (up to $t\approx 20$), choosing 
$\beta=0.1$. 

The $L_c$ evolution of an observable $\cO$ is defined by the differential 
equation
\bea
\partial_t \cO_k(z;t)&= L_c \cO_k(z;t)\quad (t\ge 0), \nonumber \\
\cO_k(z;0)&=\exp(ikz)\,;
\label{obsevol}
\eea
with
\be
L_c =\left[\partial_z-S'(z)\right]\partial_z\,.
\ee
We compare (see (\ref{fttau}) for the definition of $F_\cO$)
\bea
\int dx dy\,& P(x,y;0)\cO(x+iy;t)\ \nonumber\\
\equiv &F_\cO(t,t) \equiv \bra {\cO}(t)\ket_0
\label{lhs}
\eea
with 

\bea
\int dx\,dy\, &P(x,y;t)\cO(x+iy;0)\equiv \bra \cO(0)\ket_t\nonumber\\
&\equiv \bra \cO\ket_t\equiv F_\cO(t,0)\,.
\label{rhs}
\eea
Here $P(x,y;t)$ is the solution of the {\em real} Fokker\-Planck equation 
(FPE) 
\be
\label{realFPE}
\frac{\partial}{\partial t} P(x,y;t)= L^T P(x,y;t),
\ee
with
\be
L^T=\partial_x\left[\partial_x-K_x\right]-\partial_y K_y\,,
\ee
and initial condition (\ref{init}), which describes the time evolution of 
the probability density under the CL process. 

For our model the FPE is (\ref{realFPE}) with the drift force 
(\ref{drift}). Eq.(\ref{realFPE}) is solved numerically as well; some 
details are found in Appendix \ref{FPE}.

Fig.~\ref{corr} compares (\ref{lhs}) and (\ref{rhs}) for the Fourier 
modes 
\be
\cO_k(z)=\exp(ikz)\,.
\ee
for $k=1,2,3$, Langevin times $t$ between $0$ and $50$ and $\beta=0.1$.

It is seen that the left hand side (\ref{lhs}) reaches its asymptotic 
value already for quite short Langevin times (around $t\approx 7$). This 
is in accordance with the value of the smallest nonzero eigenvalue 
$\lambda_1\approx -1$ of $L_c$(cf. Eq.(\ref{eigenval})). For this value of 
$\beta$ also the right hand side does the same; for $t\lessapprox 20$ 
there is no difference visible between the left and the right hand sides 
(dashed and solid curves). This indicates that any boundary terms 
are negligible there.

So there is a `plateau' corresponding to the correct value in the solid 
curve, and the boundary term starts picking up around $t=20$.

\begin{figure}[ht]
\begin{center}
\includegraphics[width=1\columnwidth]{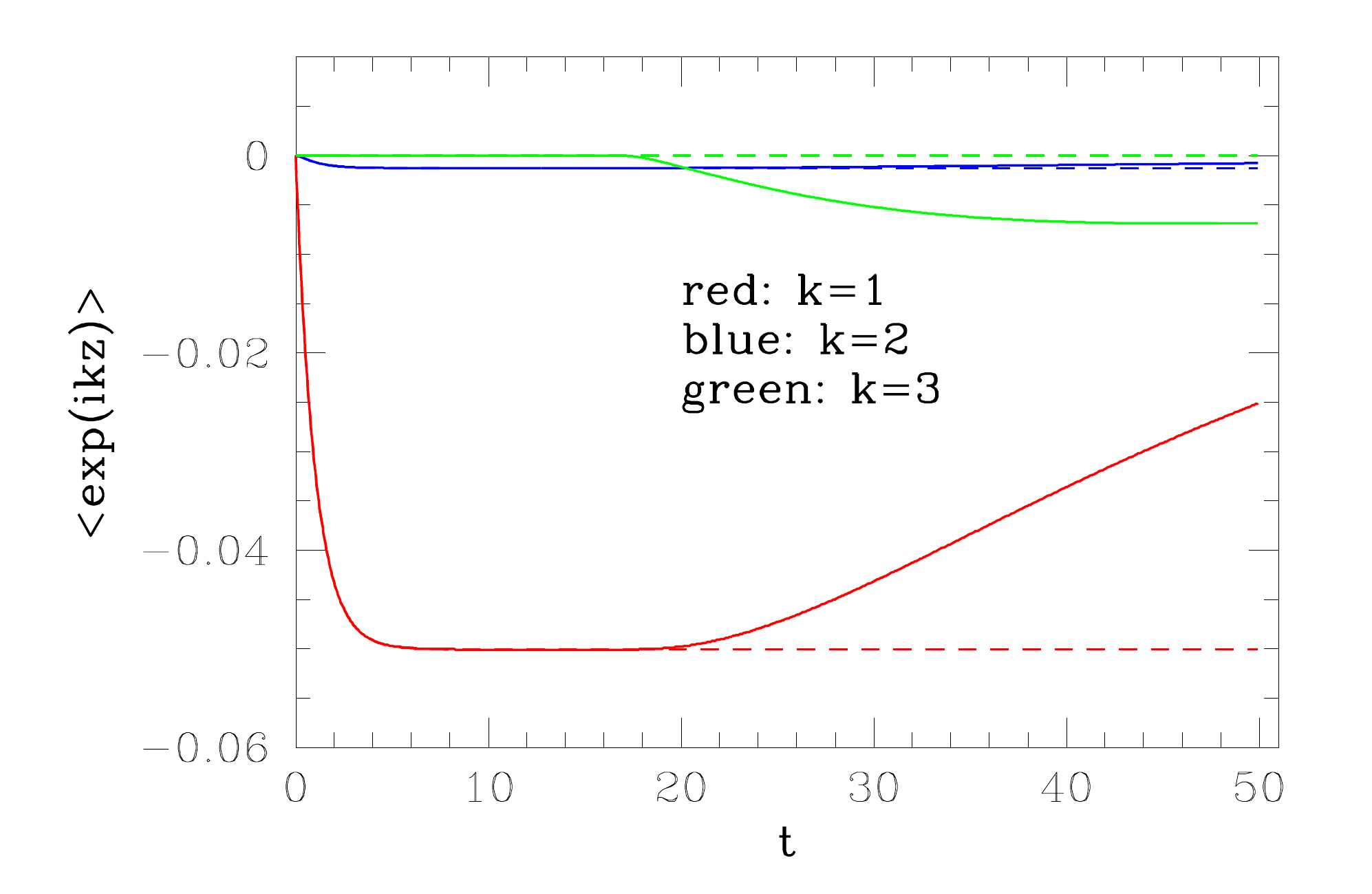}
\caption{Comparison of expectation values using the FPE evolution of $P$
(solid lines) Eq. (\ref{rhs}) with the $L_c$ evolution of the observables 
(dashed lines -- Eq. (\ref{lhs})) for $\beta=0.1$. Note that 
for times up to about $20$ the dashed and solid lines are 
practically indistinguishable.}
\label{corr}
\end{center}
\end{figure}

\begin{figure}[ht]
\begin{center}
\includegraphics[width=1\columnwidth]{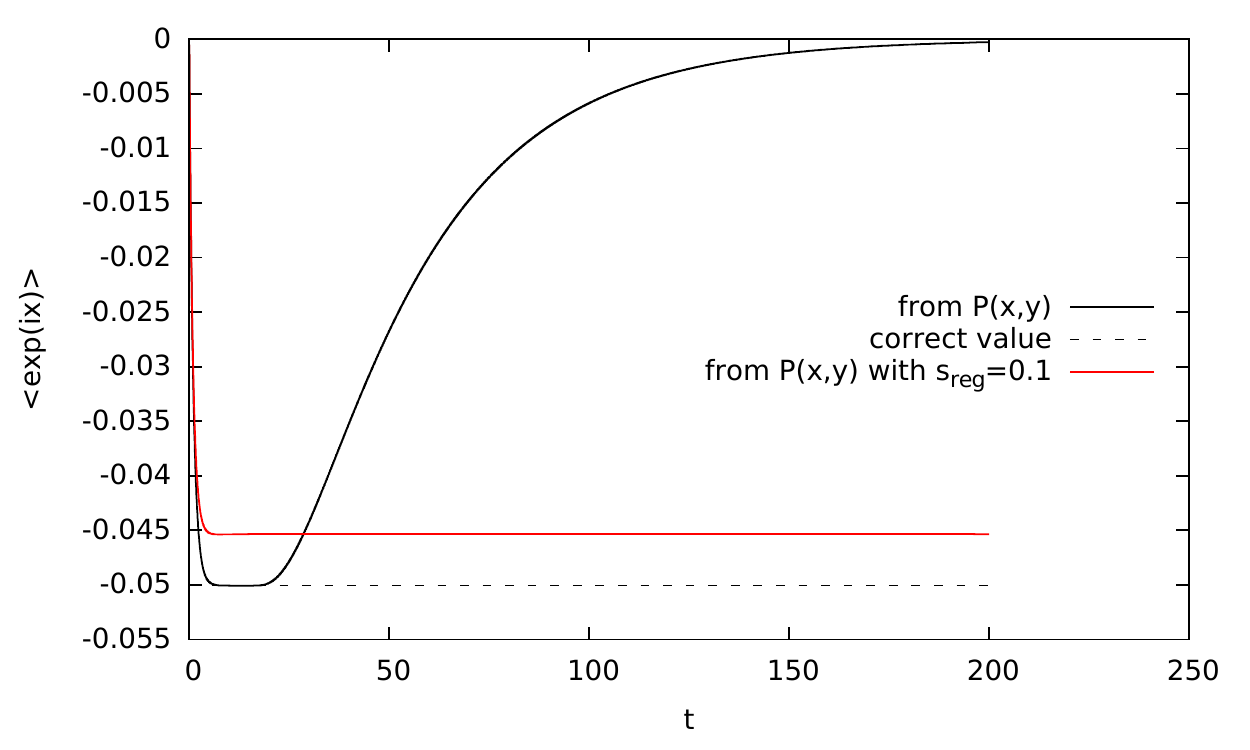}
\caption{FPE evolution of $O_1$ at $\beta=0.1$ (black). For comparison we 
show the evolution with regularization $K_{R,y} = -s\, y$, $s=0.1$ (red), 
see Section 4.}
\label{regul}
\end{center}
\end{figure}

In Fig.~\ref{regul} we show (in black) the evolution of the first mode up 
to time $t=200$. It is seen that after the plateau it converges to zero, 
the value corresponding to the stationary solution (\ref{P}) of the FPE. 
We will return to this figure in Section 5.

\subsection{Direct study of the boundary terms}

We next study explicitly the evolution of the boundary term
Eq.~(\ref{boundt}) for the modes $k=1,2,3$ and $\tau=0$ with 
Langevin time $t$. As
explained in Appendix \ref{app1} the definition of this term implies a certain
order of limits: Integrate by parts restricted to  $|y|\le Y$, send $t\to\infty$ and 
then $Y\to \infty$ (notice that this does not require a 
separate simulation but a certain processing of the data). We obtain for 
the $k$th mode in our model:

\bea
&B_k(Y;t,0)=\frac{\partial}{\partial \tau} F_k(t,\tau)
\vert_{\tau=0}\notag\\&= \beta\int_{-\pi}^{\pi} dx \sin(x)\cosh(Y)
\e^{ikx}\times\nonumber\\
&\left[P(x,Y;t)\e^{-Y}-P(x,-Y;t)\e^{Y}\right]\,.
\label{boundt1}
\eea
We first note that we can take the limit $t\to\infty$ of this expression,
using the fact that $P(x,y;t)$ indeed converges to Eq.~(\ref{P}), which 
was
verified before. We obtain
\bea
&B_k(Y;\infty,0)=  
\nonumber\\
& -2\beta\int_{-\pi}^{\pi} dx\frac{\sin(x)\cosh(Y)
\e^{ikx}\sinh(Y)}{4\pi\cosh^2(Y)}\,.
\label{bound_tanh}
\eea
For $k=\pm 1$ this can be evaluated to  
\be
B_{\mp 1}(Y;\infty,0)=\mp \frac{i\beta}{2}\tanh(Y)\,,
\ee
(converging to $\mp i\beta/2$ for $Y\to\infty$), whereas for $|k|>1$ we 
obtain $0$.

\begin{figure}[ht]
\begin{center}
\includegraphics[width=1\columnwidth]{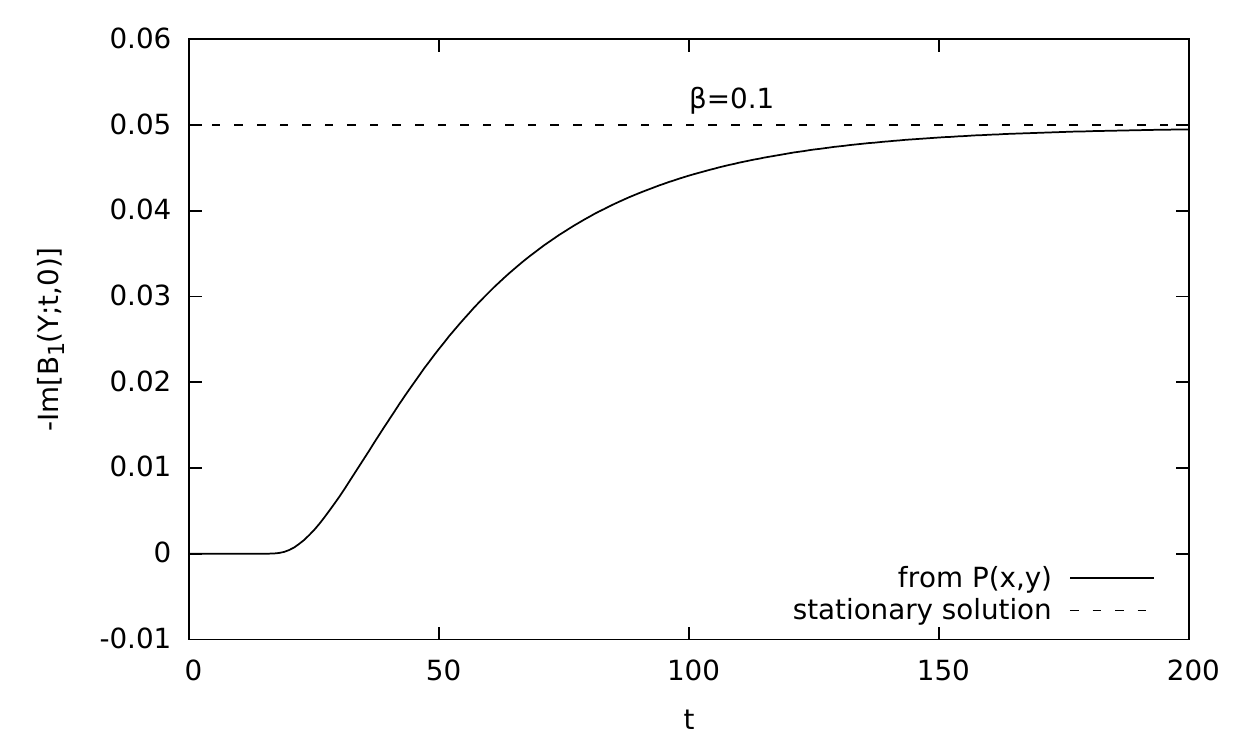}
\includegraphics[width=1\columnwidth]{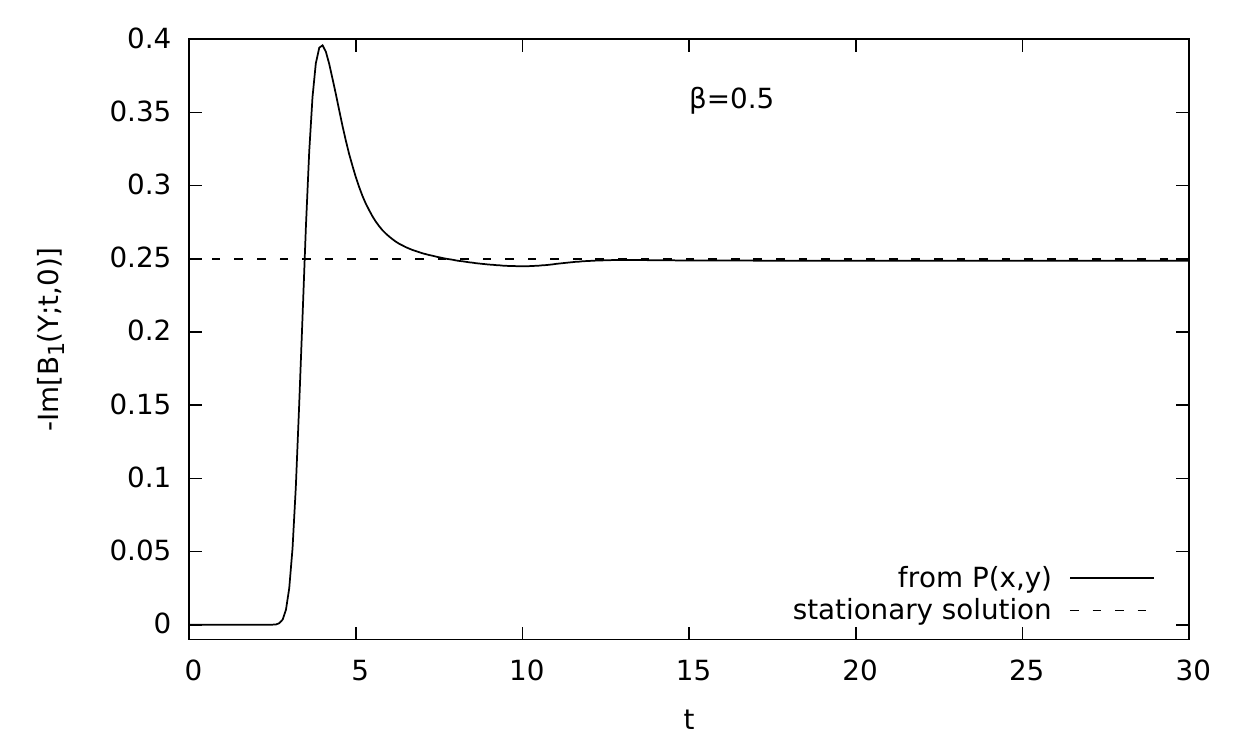}
\caption{Numerical evolution via FPE of the imaginary part of the boundary 
term Eq.~(\ref{boundt1}) for $\beta=0.1$ (top) and $\beta=0.5$ 
(bottom), $k=1$ and $Y=5$.}
\label{boundterm}
\end{center}
\end{figure}

In Fig.~\ref{boundterm} we compare $B_1$ determined numerically
for Langevin 
times up to $t=200$ with the asymptotic value at $t=\infty$ for 
$\beta=0.1$. $Y$ was chosen to be $5$ which is close to the asymptotic 
value $Y=\infty$ ($\tanh(5.)=0.99991$). We see here directly that the 
boundary term stays very small up to $t\lessapprox 20$, then picks up and 
approaches the analytically determined value $-i\beta/2$. For the value 
$\beta=0.1$ it also follows closely the difference between the first mode 
shown in Fig.~\ref{regul} and the correct value, but this cannot remain 
true for larger $\beta$.

We also checked the cases $k=2,3$ and found that $B_k$ also starts 
out very small up to about $t=20$, then increases and for large $t$ seems 
to go to the asymptotic value $0$ determined above. But one has to keep in 
mind that for $|k|\ge 2$ we are for $Y\to\infty$ evaluating a 
conditionally convergent integral; the CL process or equivalently the FPE 
evaluates that integral in a different way and may therefore produce 
different results. For $k=1$, however, there is no such subtlety and the 
boundary term $B_1(\infty;t,0)$ agrees with the slope of $F_k(t,\tau)$ at 
$\tau=0$.

\begin{figure}[ht]
\begin{center}
\includegraphics[width=1\columnwidth]{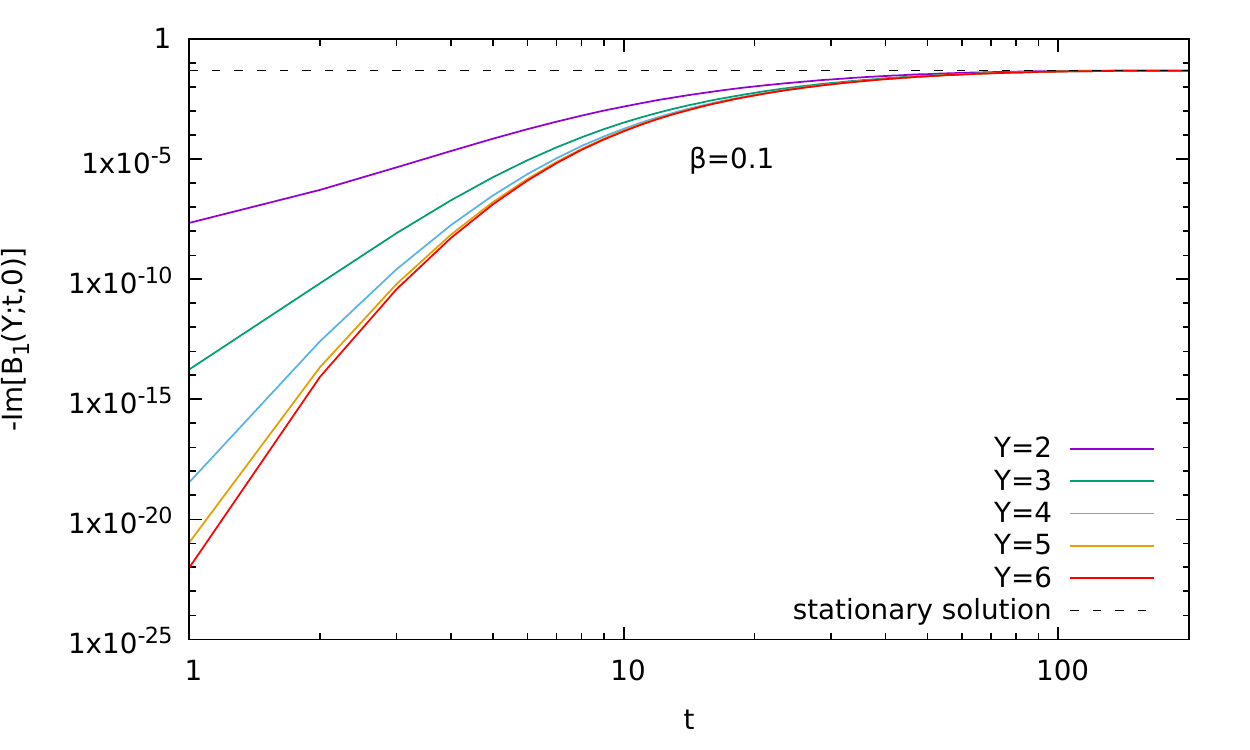}
\includegraphics[width=1\columnwidth]{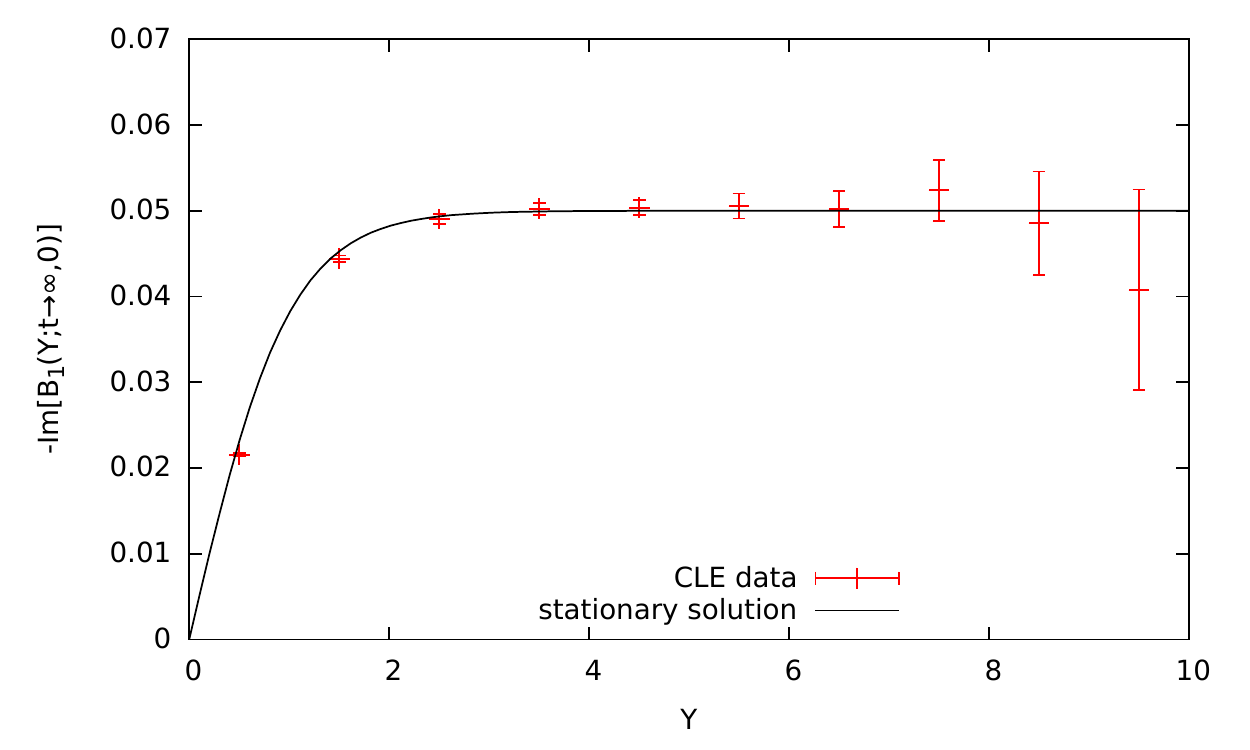}
\caption{Top panel: numerical evolution via FPE of the imaginary part of 
the boundary term Eq.~(\ref{boundt1}) for $k=1$, $\beta=0.1$ and 
different values of $Y$ vs. $t$. Bottom panel: the same boundary term 
evaluated via Langevin simulation at asymptotic $t$ vs. $Y$.}
\label{boundtermY}
\end{center}
\end{figure}

In Fig.~\ref{boundtermY} we also show the boundary term $B_1$ for 
different values of the cutoff $Y$, showing the fast approach to the 
asymptotic value. Note that in the lower panel we show the boundary term 
as measured using the CLE alone, without making use of the Fokker-Planck 
evolution, which would be prohibitively costly in a lattice model.

So we established implicitly and explicitly that boundary terms 
appear appreciably only after some Langevin time. Nonvanishing boundary 
terms at any $t>0$ invalidates the argument for correctness.

It can also seen by inspection of Eq.~(\ref{boundt1}) that the presence of 
the observable $\cO_1$ is essential; the distribution of the drift force 
alone goes to zero. Quite generally it is the product of observable, drift 
and probability $P$ that decides about the presence or absence of boundary 
terms.

\subsection{Boundary terms and skirts}
\label{bc_skirts}

Thinking now of $Y$ not as a cutoff, but as a variable, and denoting it by 
$y$ again, we see that the the first term of the boundary term 
$B_1(y;\infty,0)$, considered as a function of $y$ Eq.~(\ref{boundt1}) 
is just the probability density of the observable
\be
v(y)\equiv {\rm Im}\,\int dx K_y(x,y) \cO_1(x-iy) \sim \e^{2y} 
\ee
for large $y$. The nonvanishing of the boundary term is the fact that
\be
\lim_{y\to\infty} v(y) P_y(y)\neq 0.
\ee
On the other hand the distributions of $v$ itself has a density $p(v)$, 
related to $P$ by
\be
p(v) =  P_y(y(v))\frac{dy}{dv}\,.
\ee
This can easily be worked out, but the point is that for large $y$ 
\be
p(v)\sim v^{-2}\,,
\ee
which shows that there is no finite expectation value of $v$ since 
$v\,p(v)  $ is not integrable. 

In other words: {\em a `skirt' in the distribution of $K_y\cO$ falling off 
like the power $-2$ or more slowly corresponds to a nonvanishing (possibly 
diverging) boundary term}. Note, however, such a simple reasoning is 
only possible because here $P$ is independent of $x$.

\subsection{The interpolating function}

\begin{figure}[ht]
\begin{center}
\includegraphics[width=1\columnwidth]{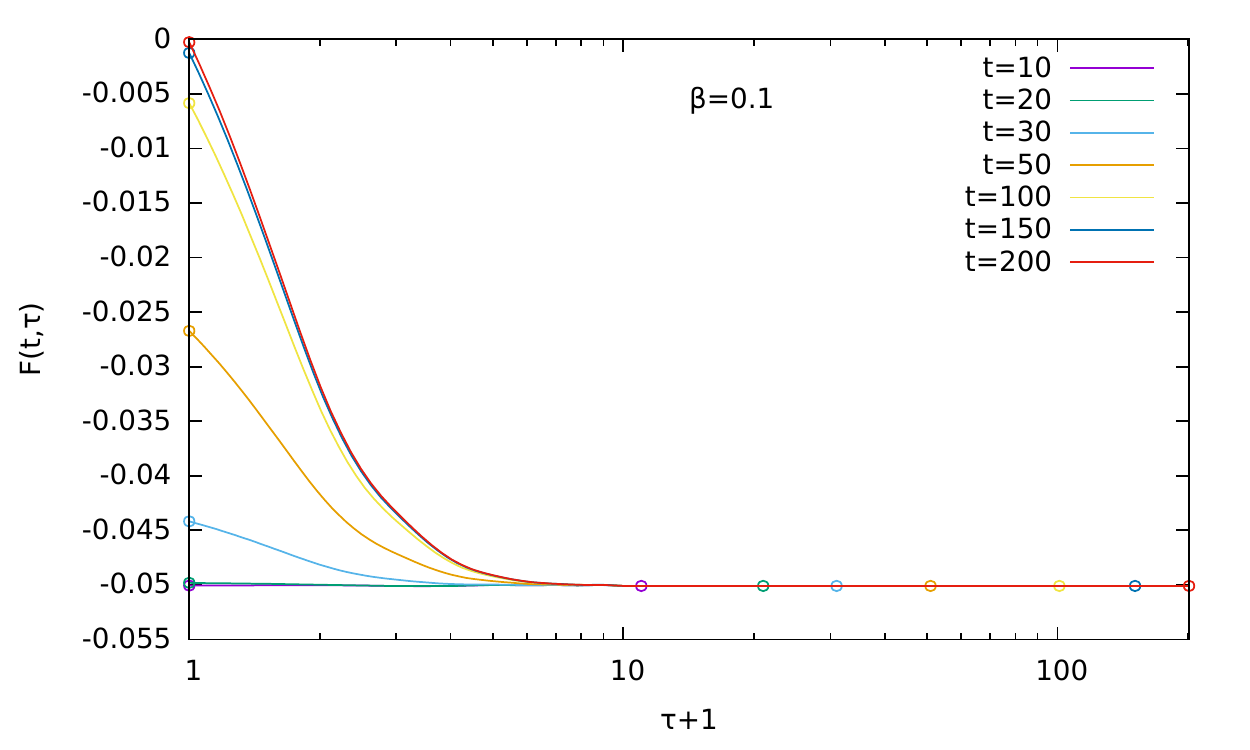}
\includegraphics[width=1\columnwidth]{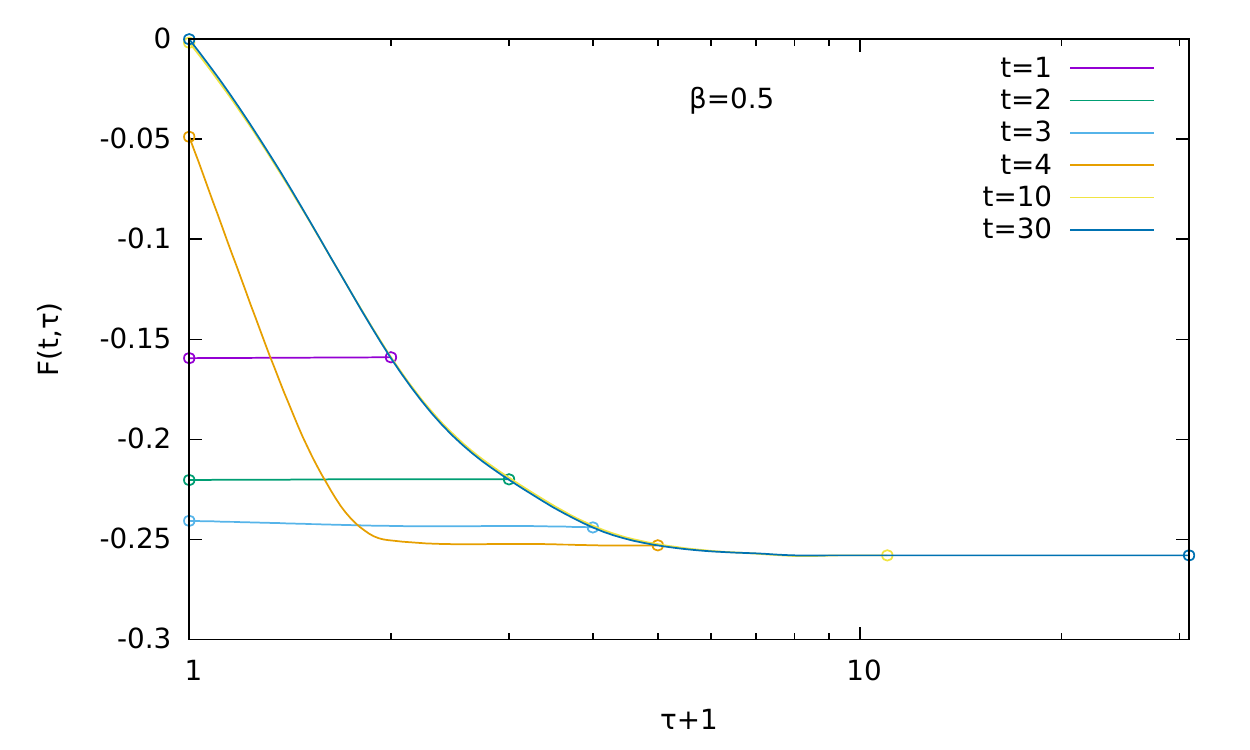}
\caption{The interpolating function $F_1(t,\tau)$ defined in (\ref{fttau})
for the first mode; $\beta=0.1$ (top) and $\beta=0.5$ (bottom) 
for various values of $t$; the small circles denote the beginning and end 
of the respective curves. 
} 
\label{fttaufig}
\end{center}
\end{figure}

So far we have only compared $F_k(t,0)$ and $F_k(t,t)$. But it is 
instructive also to look at the interpolating function $F_k(t,\tau)$
\be
F_\cO(t,\tau)\equiv \int P(x,y;t-\tau) \cO(x+iy;\tau)dxdy\,,
\label{fttau}
\ee 
which should be independent of $\tau$ for the correctness argument to 
hold. This is shown in Fig.~\ref{fttaufig} for $k=1$ for $\beta=0.1$ and
$\beta=0.5$.

Again it is seen that for $\beta=0.1$, $t\lessapprox 20$ the curves are 
flat, indicating the absence of any appreciable boundary terms. For $t>20$ 
a $\tau$ dependence develops, being maximal near $\tau=0$. This is
understandable from what we have seen: the FPE evolution of $P$ proceeds 
up to time $t-\tau$, which allows for the boundary terms to arise. On the
other hand, for $\tau\gtrapprox 7\;$ $\cO_k(z;\tau)$ has practically
reached its asymptotic limit (cf. Appendix \ref{app2}), in which only the 
constant mode survives; this constant can be pulled outside the integral 
defining $F$, so that for $t,\tau > 7 $
\be
\cO_k(z;\tau) \approx \frac{1}{2\pi}\int_{-\pi}^\pi dx'\rho(x')\cO_k(x')
= \bra \cO_k\ket_c\,
\ee
and
\be
F_k(t,\tau)\approx  \frac{1}{2\pi}\int dxdy P(x,y;t)\bra\cO_k\ket_c
= \bra \cO_k\ket_c\,.
\ee
i.e. the correct value (where we used the fact that the density $P$ is 
always normalized).

At small $t,\tau $ flat curves for $F_1(t,\tau)$ indicate that CL gives 
the correct values, however these are dependent on the initial condition 
if the process did not yet thermalize. This is seen in 
Fig.~\ref{fttaufig}, bottom plot, for $\beta=0.5$.

Notice that the slope of $F_k(t,\tau)$ appears maximal near
$\tau=0$ for large $t$. Therefore the estimation of noxious boundary terms 
as defined in (\ref{boundt1}) is relevant for judging the asymptotic 
correctness of the CL procedure -- cf. Figs.~\ref{boundterm}, 
\ref{fttaufig}.

Plots similar to Fig.~\ref{fttaufig} appeared in \cite{Aarts:2011ax} for a   
different model.

\subsection{Evolution of some marginal distributions}

For $\beta=0.1$ we saw clearly the evolution first apparently converging 
to the correct value and then departing from it (the `plateau' in 
Fig.~\ref{corr}). Similar 
behavior in Langevin time was observed in a real time $SU(2)$ lattice 
simulation \cite{Berges:2006xc}. 
This is also reflected in some marginal distributions.
 
\begin{figure}[ht]
\begin{center}
\includegraphics[width=1\columnwidth]{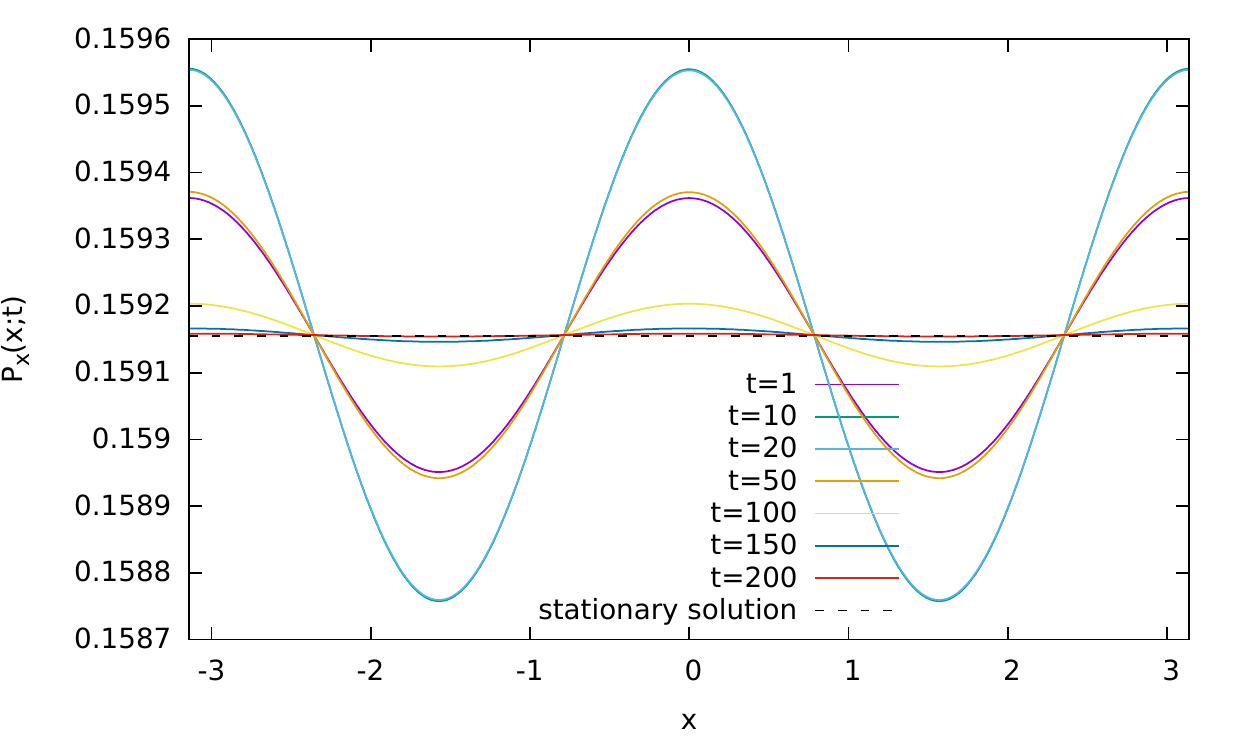}
\caption{The marginal distribution $P_x(x;t)$ obtained from solving the
Fokker-Planck equation for $\beta=0.1$.}
\label{marginalx}
\end{center}
\end{figure}

In Fig.~\ref{marginalx} we show the evolution of $P_x(x;t)=\int dy 
P(x,y;t)$ for $\beta=0.1$. It starts out flat, corresponding to our 
choice of initial condition; at $t=10$ and $t=20$ it shows maximal 
structure, while for larger $t$ it approaches a flat distribution again, 
in agreement with  (\ref{P}).

The distribution of the first mode also show a similar behavior.  
Of interest is the imaginary part. Its density is
\be
\sigma(u;t)\equiv \int dx dy P(x,y;t) \delta(\sin(x)e^{-y}-u)\,.
\label{sigmadef}
\ee
We present in  Fig.~\ref{sigmau} histograms for $\sigma(u;t)$, obtained 
from the numerical solution of the FPE; for the limiting distribution
$P(x,y;\infty)=1/(4\pi\cosh^2(y)$ we can evaluate (\ref{sigmadef})
analytically:

\begin{figure}[ht]
\begin{center}
\includegraphics[width=1\columnwidth]{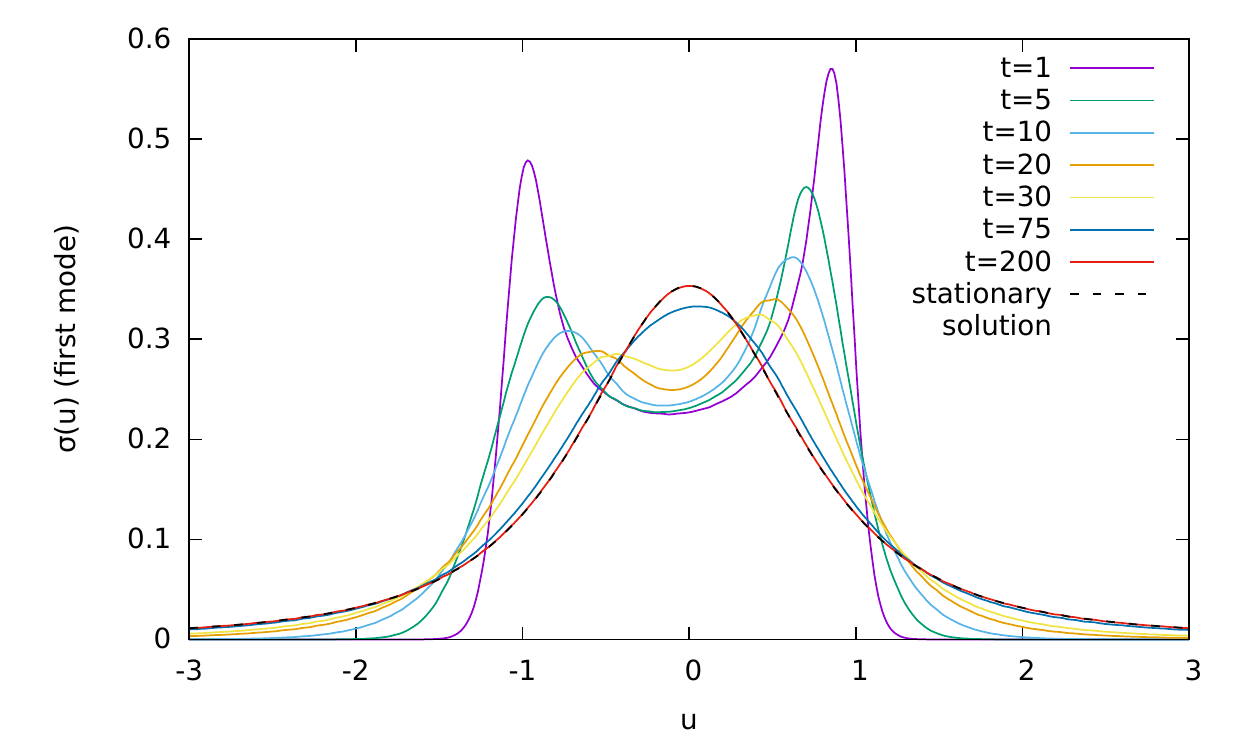}
\includegraphics[width=1\columnwidth]{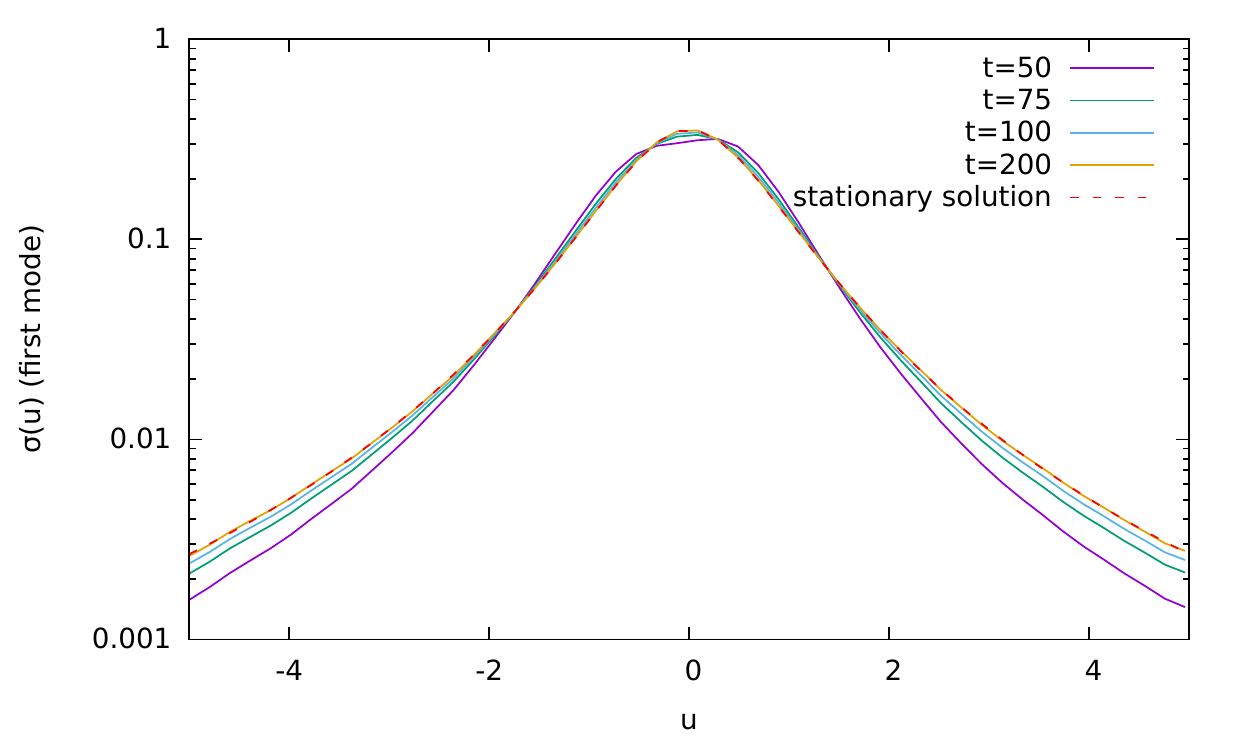}
\caption{Evolution of the distribution of the first mode $\sigma(u)$.
Top panel: times from $t=1$ to 200 in linear scale; bottom panel: times 
from $t=50$ to 200 in log scale. Again note that for $t=200$ no difference is 
visible between the numerical results and the analytic expression.} 
\label{sigmau} \end{center}
\end{figure}

\bea
\sigma(u;\infty)&=\frac{|u|}{\pi}\int_{-1}^1
dt\frac{t^2}{(t^2+u^2)^2\sqrt{1-t^2}}\nonumber\\
&=\frac{1}{2(1+u^2)^{3/2}}\,.
\label{sigma_anal}
\eea
Fig.~\ref{sigmau} shows first the development of an asymmetric structure 
with two maxima, whereas for larger $t$ one sees clearly  
the approach to the symmetric analytic result (\ref{sigma_anal}). 

\begin{figure}[ht]
\begin{center}
\includegraphics[width=1\columnwidth]{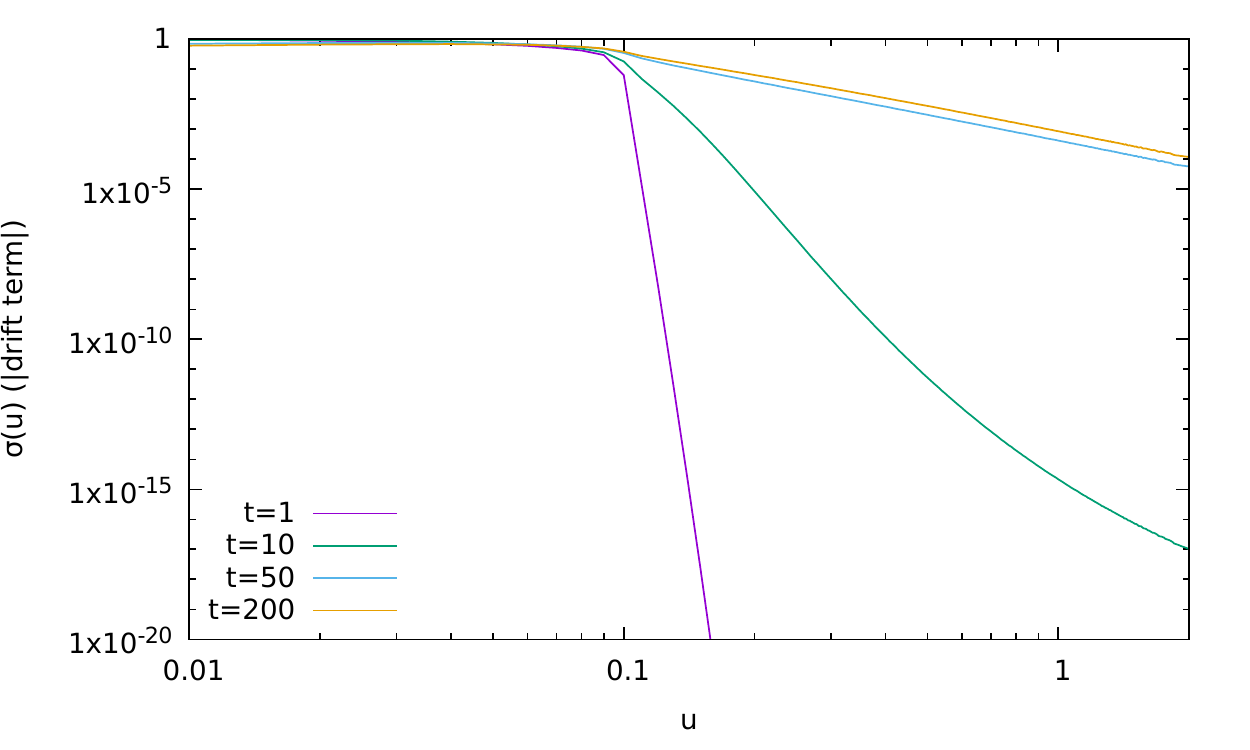}
\caption{Histograms of the drift for different Langevin times $t$ 
and $\beta=0.1$.}
\label{histodrift}
\end{center}
\end{figure}

In this context it may be of interest to compare with the criterion 
of \cite{Nagata:2016vkn}. In Fig. \ref{histodrift} we show the 
distribution of the drift itself for various Langevin times in double 
logarithmic scale. The decay always seems power-like, albeit with a very 
high power for short times. This would indicate, according to 
\cite{Nagata:2016vkn}, that even for the small times where the CL results 
seem to be correct (but not necessarily converged) there might be a tiny 
boundary term making the results incorrect by an invisible amount.

\section{Illustration of the effect of the boundary terms in a regularized
model}
\label{sect_reg}

In the preceding section we described how the boundary terms accumulate in 
the Langevin (and Fokker-Plank) evolution, spoiling the proof of 
convergence such that the process would lead to wrong results. 

Here we want to explicitly see the effect of those terms by considering a 
`regularization' of the model using a damping term in the action, 
$S_R = \frac{s}{2}\,x^2$, 
which leads to a modification of the drift by $K_R(z)=- s\,z$ (a similar 
regularization has been used in \cite{Loheac:2017yar}; we thank
 J. Drut and A. C. Loheac for making us aware of this).
The philosophy of this regularization is very similar to that of dynamical 
stabilization \cite{Attanasio:2018rtq}. In both cases, and different from 
modifications using symmetries, such as in the gauge cooling paradigm, 
the dynamics is really changed, but in a way intended to be controllable.  

For $s=0$ we regain the original model Eq.~(\ref{model}) (including its 
problems) while for $s > 0$ we should observe an interplay between the 
original tendency to build boundary terms and their damping in the 
modified model, allowing us to estimate the effect of these terms. This 
particular modification leads to loss of periodicity in $x$ which becomes 
noncompact at $s>0$. 
The CLE process was allowed to drift unbounded in the full $z$ plane
and the exact integral was correspondingly done in the infinite interval. 
The following plots show $O_1= \Im \bra e^{iz} \ket$. 
\begin{figure}[ht]
\begin{center}
\includegraphics[width=1\columnwidth]{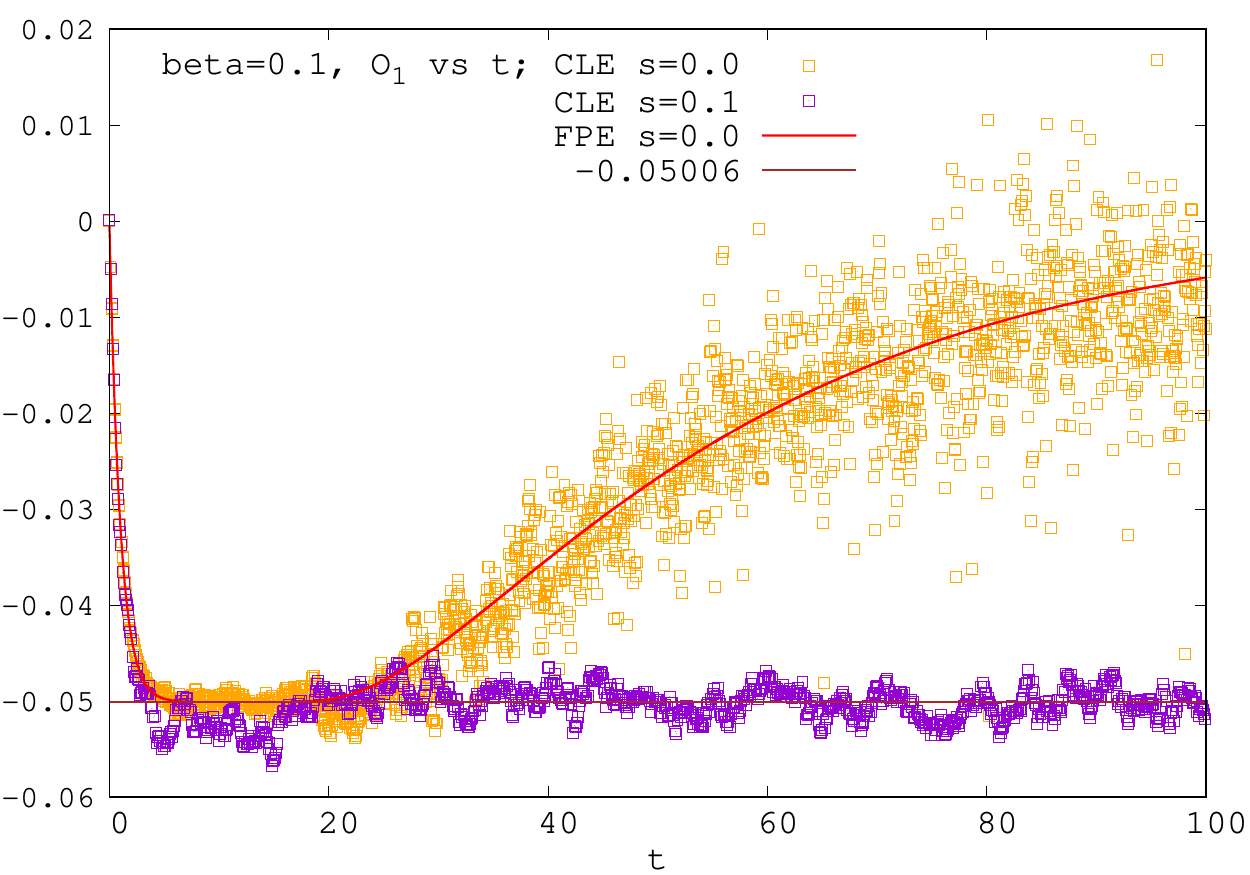}
\includegraphics[width=1\columnwidth]{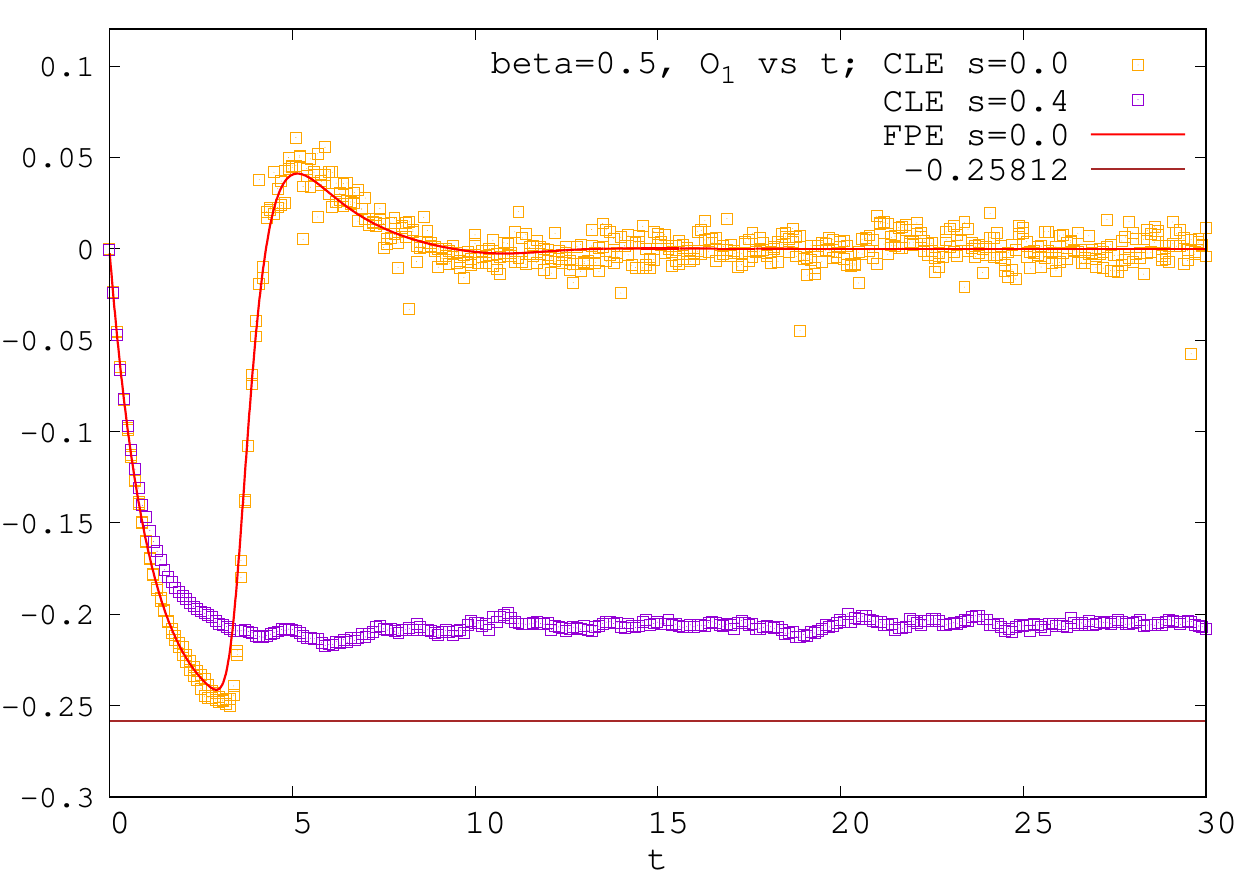}
\caption{Comparison of the  $O_1$ expectation values from
FPE (solid line) and CLE  and from regularized CLE, vs $t$. {Top}:
$\beta=0.1,\, s=0$ and $0.1$, respectively. { Bottom}: 
$\beta=0.5,\, s=0$ and  $0.4$,  respectively.}
\label{f.regt}
\end{center}
\end{figure}
We see from Fig.~\ref{f.regt} that the regularization stabilizes the 
expectation values in the CLE evolution. When the non-regularized data 
show a plateau at the correct value for intermediary $t$ the 
regularization extends this plateau into the asymptotic region 
($\beta=0.1$ case). When a plateau is missing the regularization still 
stabilises the expectation value (EV) but at a value shifted from 
the correct one ($\beta=0.5$ case), since now a larger $s$ is needed to 
counteract the boundary terms.
 
Note that an alternative regularization of the process itself is to modify 
{\it only the imaginary drift} by a damping term $K_{R,y}=-s\,y$. This leads to 
similar results (see Fig.~\ref{regul}, here from the FPE evolution), and 
has the advantage that periodicity in $x$ is preserved. We preferred the 
action variant, however, also since it allows us to obtain exact correct
results for the regularized model by simple numerical integration.

\begin{figure}[ht]
\begin{center}
\includegraphics[width=1\columnwidth]{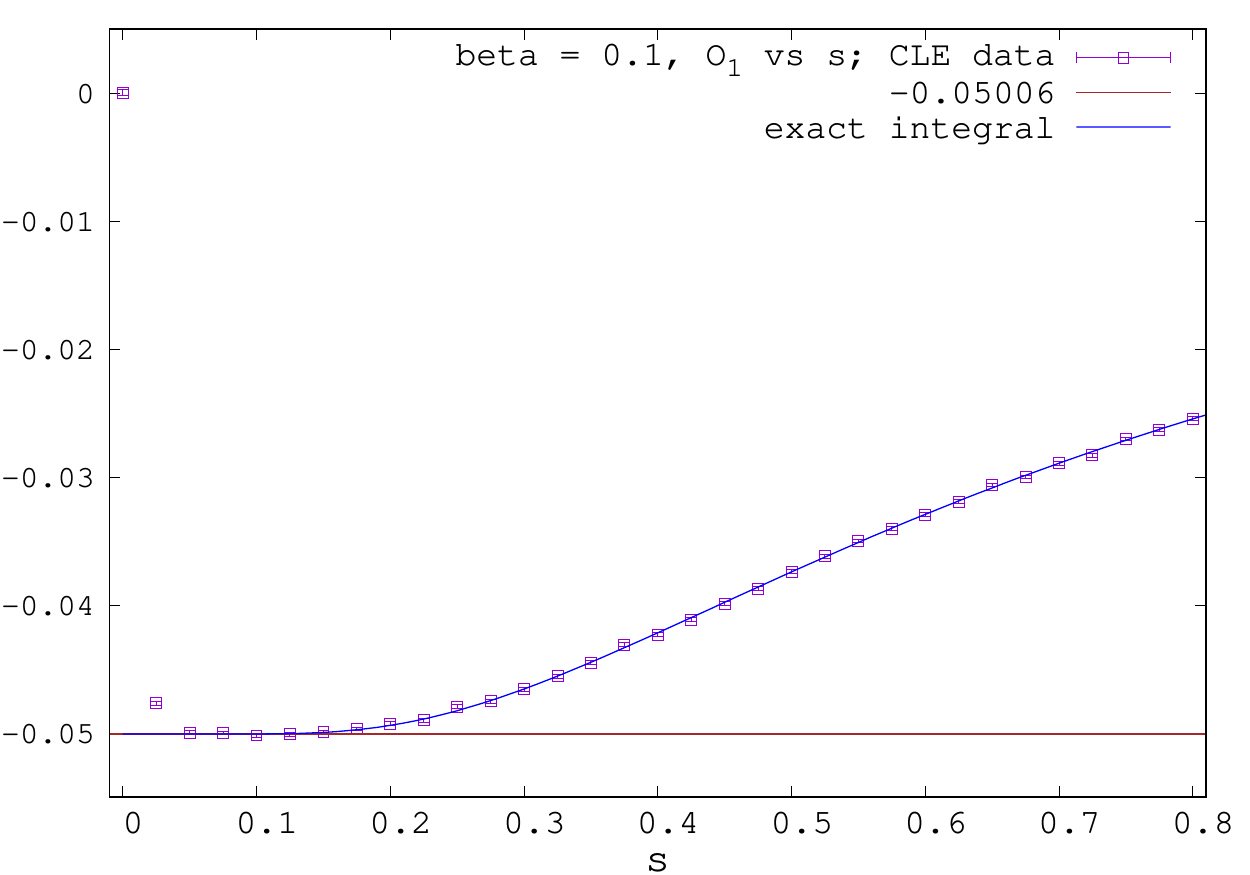}
\includegraphics[width=1\columnwidth]{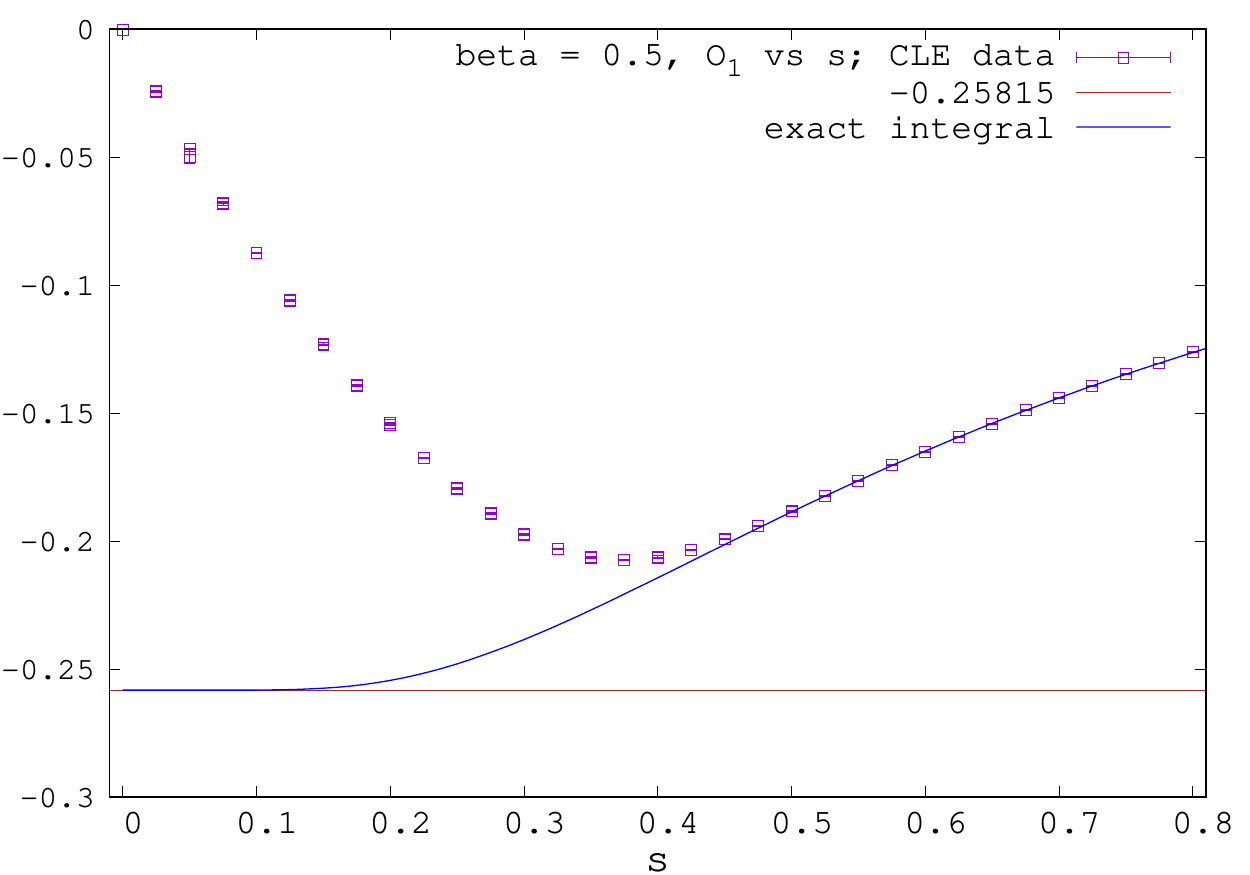}
\caption{Dependence of $O_1$ on the regularization parameter $s$ of the CLE
 simulation  at $\beta =0.1$ (top) and $0.5$ (bottom). The solid lines 
show the exact correct values from numerical integration.}
\label{f.regs}
\end{center}
\end{figure}

In Fig.~\ref{f.regs} we show the $s$ dependence in CLE for the regularized 
model for the same values of $\beta$. The plots suggest an extrapolation 
toward the exact expectation value (EV) for $s \rightarrow 0$, however 
this might not be simply linear, but depend on the particular 
regularization, $\beta$, etc. Therefore we mean this discussion not yet 
as a direct cure but mainly as illustration of the effects of the boundary 
terms on the EV's. For $k=1$, e.g., these effects can be estimated from 
the distance between the CLE regularized data and the exact values from 
the numerical integration of the regularized model: As can be seen from 
the figure at small $s$ these effects are still present while gradually 
vanishing with increasing $s$.
\begin{figure}[ht]
\begin{center}
\includegraphics[width=1\columnwidth]{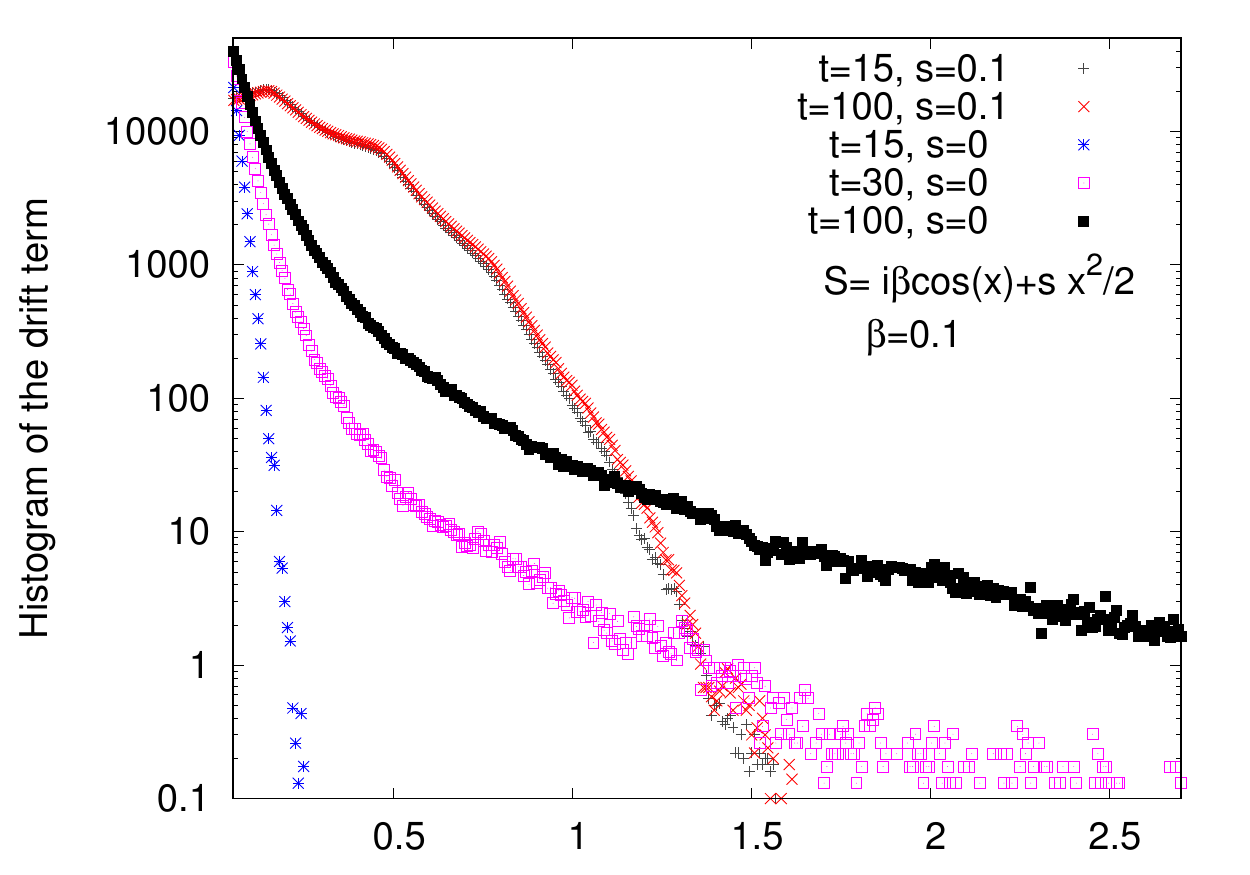}
\caption{Histogram of the drift for $\beta=0.1$ and various Langevin 
times, comparing $s=0$ and $s=0.1$.}
\label{histodrifts}
\end{center}
\end{figure}

Finally it is again instructive to look at the histograms of the drift 
itself, as advocated by \cite{Nagata:2016vkn}, to see the effect of the 
regularization.  This is shown in Fig.~\ref{histodrifts}. One can see that 
the distribution seems to show power-like decay $s=0$ whereas for $s=0.1$ 
the decay appears to be exponential. This supports the criterion of 
\cite{Nagata:2016vkn}, because for the value of $\beta=0.1$ used here, 
already $s=0.1$ suffices to bring the CL results into agreement with the 
correct results of the regularized as well as the unregularized model, 
which are indistinguishable in this case, as shown in Fig.~\ref{f.regs}.

\section{Conclusions}

We have in great detail analyzed a simple example in which the CL fails, 
establishing very explicitly that the failure is due to boundary terms 
spoiling the correctness argument, as argued already long ago 
\cite{Aarts:2009uq,Aarts:2011ax}. The absence of such boundary terms 
requires that the product of observable, drift force and probability 
distribution ($\cO K P$) goes to zero in the noncompact (imaginary) 
directions. The relation between boundary terms and `skirts', i.~e. decay 
of distributions was addressed in Subsection~\ref{bc_skirts}, making clear 
that possible skirts in the distribution of the product $\cO K P$ and not 
just $KP$ are relevant. Remarkably, the criterion proposed by 
\cite{Nagata:2016vkn} does not involve the observable. 

Generally the $\tau-$dependent boundary term (\ref{boundtgeneral}) cannot 
be estimated in a realistic (lattice) calculation. Fortunately, however, 
the considerations in this paper suggest that relevant for the correctness 
of the asymptotic (large $t$) EV's is the boundary term at $\tau=0$, 
$B_k(t,0)$, as defined in (\ref{boundt1}). This term appears to 
approximately maximize $B_k(t,\tau)$ and it stabilizes at large 
$t$; it is accessible in principle to online monitoring using the CLE 
alone, and may provide a correctness criterion for the EV's obtained in 
the CL simulation.

As remarked before, uncovering the boundary term requires a certain 
processing of the data obtained in the simulation, implying essentially 
sampling first at a fixed value of a quantity specifying the boundary in 
the non-compact directions (in a lattice gauge theory for instance the 
unitarity norm or some other related quantity) before taking the other 
limits. This however does not require a separate simulation.

{\em Acknowledgments:} M.~S., E.~S. and I.-O.~S. gratefully acknowledge 
kind support from DFG under Grant Sta 283/16-2. D.~S. gratefully 
acknowledges funding by the DFG grant Heisenberg Programme (SE 2466/1-2).
The authors acknowledge support by the High Performance and 
Cloud Computing Group at the Zentrum f\"ur Datenverarbeitung of the 
University of T\"ubingen, the state of Baden-W\"urttemberg through bwHPC
and the German Research Foundation (DFG) through grant no INST 37/935-1 FUGG.

\appendix
\section{The argument for correctness revisited}
\label{app1}
   
We briefly revisit the formal proof of correctness for the CL method for 
our simple periodic one-dimensional models, spelling out the conditions 
needed for it to work as well as the mechanisms that may lead either to 
no convergence or `wrong convergence' of the CL process (cf. 
\cite{Aarts:2011ax,Aarts:2017vrv,Seiler:2017wvd}).

$P(x,y;t)$ is the time dependent probability distribution corresponding to 
the CL process, determined by the {\em real} Fokker-Planck equation 
(\ref{realFPE}).

We also consider the time evolution of the complex density $\rho(x;t)$
by the {\em complex} Fokker-Planck equation
\be
\label{complexFPE}
\frac{\partial}{\partial t} \rho(x;t)= L_c^T \rho(x;t),
\ee
where now the complex Fokker-Planck operator $L_c^T$ is
\be
\label{fpc0}
L_c^T =  \partial_x \left[\partial_x+S'(x)\right]\,.
\ee
The initial conditions for (\ref{realFPE}) and (\ref{complexFPE}) are 
required to be consistent, i.e.
\bea
P(x,y;0)&=\rho(x;0)\delta(y)\ge 0\nonumber \\
 \int\rho(x;0)dx&=1\,.
\eea
The crucial point is that one can now, under conditions to be spelled out
below, show that 
\be
\int dx\cO(x) \rho(x;t)= \int dx dy \cO(x+iy) P(x,y;t)\,.
\label{correctness}
\ee
If in addition the operator $L^T_c$ has spectrum in the left half plane
with 0 a nondegenerate eigenvalue, if follows that
\be
\lim_{t\to\infty}\int dx\cO(x) \rho(x;t)= \int dx\cO(x) \rho(x)\,
\ee
and by (\ref{correctness})
\bea
\lim_{t\to\infty} \int dx dy &\cO(x+iy) P(x,y;t)\\
&=\int dx\cO(x)
\rho(x)\,.
\eea
By our choice of initial conditions, (\ref{correctness}) holds for
$t=0$. For $t>0$ we consider $F_\cO(t,\tau)$ defined in Eq.~(\ref{fttau}), 
which interpolates between the two sides of (\ref{correctness}):
\be
F_\cO(t,\tau)\equiv \int P(x,y;t-\tau) \cO(x+iy;\tau)dxdy\,,
\ee
with $\cO(x+iy;t)$ defined by solving the differential equation
(\ref{obsevol})
$L_c$, the complex Langevin operator, is the transpose of $L^T_c$:
\be
L_c =\left[\partial_z-S'(z)\right]\partial_z\,.   
\ee
We call the solution of Eq.~(\ref{obsevol}) the `$L_c$  evolved' 
observable.

The interpolating property follows from

\bea
F_\cO(t,0)&= \int dx dy \cO(x+iy) P(x,y;t)= \bra \cO\ket_t \notag\\ 
F_\cO(t,t)&=\int dx\cO(x,t) \rho(x;0)= \bra \cO(t)\ket_c\,,
\eea
see Section~\ref{indirect}, Eqs.~(\ref{lhs}) and (\ref{rhs}). The first 
equality is obvious, the second one follows by integration by parts in 
$x$; because of periodicity there are no boundary terms. 
(\ref{correctness}) would follow if we could prove
\bea
\label{correctinter}
&\frac{\partial}{\partial \tau} F_\cO(t,\tau) =\nonumber\\ &
-\int \left(L^T P(x,y;t-\tau)\right)\cO(x+iy;\tau)dxdy\notag\\ 
& + \int  P(x,y;t-\tau) L_c\cO(x+iy;\tau) dxdy=0\,.
\eea
This would again follow from integration by parts, provided there 
are no boundary terms. For the term $\partial_x^2$ of both $L^T$ and $L_c$ 
this is obvious because of periodicity, so we can drop these terms, 
obtaining
\begin{align}
&\frac{\partial}{\partial \tau} F_\cO(t,\tau) =\nonumber\\
&\int \cO(x+iy;\tau)(\partial_x K_x +\partial_y K_y) P(x,y;t-\tau)
dxdy\notag\\
& - \int  P(x,y;t-\tau) S'(x+iy)\partial_x\cO(x+iy;\tau) dx dy\,.
\label{diff1}
\end{align}
In \cite{Aarts:2009uq} we argued that $\cO(x+iy;\tau)$ is holomorphic
for any $\tau$, i.e. it obeys the Cauchy Riemann equations
\be
\partial_y \cO(x+iy;\tau)= i \partial_x \cO(x+iy;\tau)\,.
\ee
This allows us to write the second term of the right hand side of 
(\ref{diff1}) as
\be
\int  P(x,y;t-\tau) (K_x\partial_x+K_y \partial_y)
\cO(x+iy;\tau) dx dy\,.
\label{diff0}   
\ee
Again the part involving $\partial_x$ can be canceled against the
corresponding term in the first term of (\ref{diff0}) using integration
by parts in $x$, so we only have to consider
\begin{align}
&\frac{\partial}{\partial \tau} F_\cO(t,\tau) =\nonumber\\
&\int \left(\partial_y K_y P(x,y;t-\tau)\right)
\cO(x+iy;\tau)dxdy\notag\\
& +\int  P(x,y;t-\tau) K_y \partial_y\cO(x+iy;\tau) dx dy\,.
\label{diff2}
\end{align}
We have to interprete this as a the limit $Y\to\infty$ of the integral
restricted to $|y|\le Y$. For finite $Y$ (\ref{diff2}), 
since the integrand is a total derivative, this is given by the 
boundary term
\bea
&B_\cO(Y;t,\tau)\equiv\nonumber\\
&\int [K_y(x,Y)P(x,Y;t-\tau)\cO(x+iY;\tau)\notag\\
&-K_y(x,-Y)P(x,-Y;t-\tau)\cO(x-iY;\tau) ]dx\,.
\label{boundtgeneral}
\eea
Evaluating this term at $\tau =0$ leads then for our model to 
(\ref{boundt1}) where we can then take the large $t$ limit to obtain for 
our model, obtaining (\ref{bound_tanh}). This form of the boundary 
term makes clear that correctness requires sufficient decay of the 
products $K_y P\cO$.

Notice that if we take the $t \rightarrow \infty$ limit directly in
(\ref{correctinter}) the first term vanishes by stationarity 
and the second one 
leads to the `Correctness Conditions' (CC) defined in \cite{Aarts:2011ax} 
and is approximately zero by stationarity. Hence, it might appear 
that the boundary term vanishes and we might erroneously conclude 
correctness of the results. Therefore the CC, while expressing 
convergence and being necessary for correctness, are not sufficient.

\section{The correct evolution}
\label{app2}

What was called the `correct time evolution' $\bra \cO(t)\ket_0 $ 
of the expectation value of $\cO$ is simply the expectation value in the 
starting probability density $P(x,y;0)$ of the $L_c$ evolved observable 
$\cO$, see Eq.(\ref{lhs}). To analyze this we rewrite the Langevin 
operator $L_c$ in the basis of Fourier modes:
\bea
L_c \exp(ikx) &= -k^2 \exp(ikx) \nonumber\\&-\frac{i\beta}{2} k \exp(i(k+1)x)\nonumber\\&+
\frac{i\beta}{2} k \exp(i(k-1)x)\,
\eea
or equivalently, for a general observable
\be
\cO(x)=\sum_k a_k \exp(ikx)
\ee
\bea
(L_c a)_k=&-k^2 a_k - \frac{i\beta}{2} (k-1) a_{k-1} \nonumber\\+& \frac{i\beta}{2}
(k+1)a_{k+1}\,.
\label{obsevol2}
\eea
So $L_c$ is represented on the Fourier coefficients by the sparse
infinite matrix with elements
\bea
(L_c)_{kl}=&-k^2\delta_{kl}-\frac{i\beta}{2} (k-1)\delta_{k-1,l}\nonumber\\
&+
\frac{i\beta}{2} (k+1)\delta_{k+1,l}\,.
\eea
It is easy to compute numerically the action of $\exp(t L_c)$ on  
observables of the form $\cO_k=\exp(ikx)$; cutting off the modes at
$|k|\ge K$ with $K=50$ and $K=150$, and for $t=50$, gave identical
results, with only the constant mode surviving. Its value agrees to at 
least 5 digits with
\be
\lim_{t\to\infty} \exp(t L_c) \cO_k= \int dx \rho(x)\cO_k(x)\,
\ee
i.e. the correct expectation value.

We also checked, using Mathematica, that the eigenvalues of the   
truncated matrix $(L_c)_{kl}$ have negative real part except for the
unique zero eigenvalue corresponding to $a_k\propto\delta_{k0}$. All
nonzero eigenvalues are real and doubly degenerate. The one with the
smallest modulus determines the approach to the infinite time limit; it
depends only weakly on $\beta$, e.~g.
\bea
&\lambda_1=-0.998333\;(\beta=0.1)\,;  \nonumber\\ &\lambda_1= -0.832189\;
(\beta=1)
\label{eigenval}
\eea
{\it Remark:} It is easy to show that by a similarity transformation
$L_c$ can be transformed into the {\em dissipative} operator
\bea
-H &= \exp(S/2) L \exp(-S/2)\nonumber\\
&= \frac{d^2}{dx^2} -\beta^2
\sin^2(x)-\frac{i\beta}{2}\cos(x)\,.
\eea
Dissipativity means $-H-H^*\le 0$, which is obvious.  For such operators
general theorems guarantee that the spectrum is contained in the left half
of the complex plane (see for instance \cite{davies}). It is also not
hard to see that there is exactly one vector with eigenvalue zero.

\section{Remarks on the numerical solution of the FPE}
\label{FPE}

The real Fokker-Planck equation in our case is
\begin{align}
&\frac{\partial P(x,y;t)}{\partial t} =
\left[\partial_{x}\left(\partial_x - K_x\right) - \partial_y K_y\right]
P(x,y;t) \nonumber \\ =
& [\partial_{x}^2 + \beta(-2\sin x \sinh
y + \cos x \sinh y \partial_{x} \nonumber\\
&- \sin x \cosh y
\partial_{y})]P(x,y;t)\, .
\label{modelFPE}
\end{align}

Discretizing \eqref{modelFPE} in $x$ and $y$ using symmetric derivatives 
yields
\begin{align}
&P(x, y; t+dt) =\nonumber\\
& \frac{1}{dx^2} \left(P(x_+, y; t) -2P(x,y;t) + 
P(x_-,y;t)\right)\nonumber\\
-& 2\beta \sin x \sinh y P(x,y;t)\nonumber \\
+& \frac{\beta}{2dx} \cos x \sinh y 
\left(P(x_+,y;t)-P(x_-,y;t)\right)\nonumber \\
-& \frac{\beta}{2dy} \sin x \cosh y \left(P(x,y_+;t)-P(x,y_-;t)\right)\, ,
\end{align}
where we defined $x_{\pm}=x\pm dx$ and similarly for $y$.
In case of a regularization term in the $y$-drift  $K_y\rightarrow K_y-s_y 
y$ (see Section~\ref{sect_reg}), additional terms occur
\begin{align}
&P(x, y; t+dt) \rightarrow  P(x, y; t+dt) + s_y P(x,y;t) \nonumber \\
&+ s_y \frac{y}{2 dy}\left(P(x,y_+;t)-P(x,y_-;t)\right)\, .
\end{align}

We solved the Fokker-Planck equation on an $x$-$y$-grid with parameters 
$dt=10^{-6}$, $dx=0.005=dy$, a cutoff in $y$-direction of $Y=5$ was found 
to be sufficient (compare \eqref{bound_tanh}, 
$\tanh(5)\approx\tanh(\infty)$), and a cutoff in $x$-direction of 
$X=3.14$, which is due to the $2\pi$ periodicity of the problem. Boundary 
conditions in $x$ and $y$ were both chosen to be periodic.
Initial condition were chosen according to \eqref{init}, however the 
$\delta$-function was smeared out slightly to avoid numerical issues; so 
we 
actually used 
\begin{equation}
P(x,y;0)=\frac{1}{2\pi\sqrt{2\pi \sigma_{y}^2}}e^{-\frac{y^2}{2\sigma_y^2}}\, ,
\end{equation}
where we chose $\sigma_y=0.1$.
Note that using this discretization it is hard to resolve the higher 
modes. This can be done more easily when solving the Fokker-Planck 
equation in Fourier space, where it is given by
\begin{align}
&P(k,y;t+dt)= -k^2P(k,y;t) \nonumber\\
 &-\frac{i\beta}{2}\sinh(y)\left(k_- P(k_+,y;t)+ 
k_+P(k_-,y;t)\right)\nonumber\\
&+\frac{i\beta}{4dy}\cosh(y)\left(P(k_+,y_+;t)-P(k_-,y_+;t)\right)\nonumber\\
&+\frac{i\beta}{4dy}\cosh(y)\left(-P(k_+,y_-;t)+P(k_-,y_-;t)\right)\, ,
\end{align}
where $k_\pm=k\pm 1$ and similarly for $y$.
Here we chose $dt=0.5\times 10^{-5}$, $k\in\{-19,\ldots,20\}$, $dy=\sqrt{dt}$, $Y\approx 2.8$ with antiperiodic boundary conditions in $k$ for the imaginary part of $P(k,y;t)$ and periodic boundary conditions for the real part in $k$ and for $y$.
After $t\sim 30$ or so the result strongly depends on the choice of 
discretization. Hence, we use the $k$-$y$ discretization to resolve the 
plateaus in the higher modes and the $x$-$y$-discretization for everything 
else.

The solution to the Fokker-Planck equation shows that for $\beta=0.1$ the 
evolution initially follows the correct evolution. This is suggested by 
Figs.~\ref{fttaufig} and \ref{boundterm}. By looking at $P(x,y;t)$ in 
Figs.~\ref{marginal} and \ref{marginalx} and the histogram of the first 
mode in Fig.~\ref{sigmau}, one can see that initially nontrivial 
structures occur. Those die out and everything approaches the asymptotic 
solution, which yields the wrong results. This strengthens the argument 
that until $t\sim 20$ or so CLE yields the correct solution but then the 
occurrence of boundary terms leads to wrong convergence.


\end{document}